\documentclass[english,prl,aps,showpacs,twocolumn]{revtex4-1}

\def\be{\begin{equation}}
\def\ee{\end{equation}}
\usepackage{changes}
\usepackage{graphicx}
\usepackage{bm}
\usepackage[T1]{fontenc}
\usepackage[latin9]{inputenc}
\usepackage{amsmath}
\usepackage{amssymb}
\usepackage{float}
\usepackage[bookmarks=true,
   colorlinks=true,
   linkcolor=blue,
   urlcolor=blue,
   citecolor=blue,
   bookmarks=true,
   hyperindex=true
]{hyperref}
\usepackage{natbib}

\begin{document}
\title{Scale invariance of a spherical unitary Fermi gas}

\author{Lu Wang$^{1,2}$}
\thanks{These authors contributed equally to this work}

\author{Xiangchuan Yan$^{1}$}
\thanks{These authors contributed equally to this work}

\author{Jing Min$^{1,2}$}
\thanks{These authors contributed equally to this work}

\author{Dali Sun$^{1}$}
\thanks{Corresponding author: dlsun@wipm.ac.cn}

\author{Xin Xie$^{1,2}$, Shi-Guo Peng$^{1}$}

\author{Mingsheng Zhan$^{1}$}

\author{Kaijun Jiang$^{1}$}
\thanks{Corresponding author: kjjiang@wipm.ac.cn}

\affiliation{$^{1}$State Key Laboratory of Magnetic Resonance and Atomic and Molecular Physics, Innovation Academy for Precision Measurement Science and Technology, Chinese Academy of Sciences, Wuhan 430071, China}
\affiliation{$^{2}$University of Chinese Academy of Sciences, Beijing 100049, China}

\date{\today}

\begin{abstract}
A unitary Fermi gas in an isotropic harmonic trap is predicted to show scale and conformal symmetry that have important consequences in its thermodynamic and dynamical properties. By experimentally realizing a unitary Fermi gas in an isotropic harmonic trap, we demonstrate its universal expansion dynamics along each direction and at different temperatures. We show that as a consequence of SO(2,1) symmetry, the measured release energy is equal to that of the trapping energy. We further observe the breathing mode with an oscillation frequency twice the trapping frequency and a small damping rate, providing the evidence of SO(2,1) symmetry. In addition, away from resonance when scale invariance is broken, we determine the effective exponent $\gamma$ that relates the chemical potential and average density along the BEC-BCS crossover, which qualitatively agrees with the mean field predictions. This work opens the possibility of studying non-equilibrium dynamics in a conformal invariant system in the future.
\end{abstract}

\maketitle

Strongly interacting Fermi gases are created by tuning the interaction strength between atoms of different spin states via Feshbach resonance \cite{Chin2010, OHara2002}. The unitary Fermi gas, realized when the $s$-wave scattering length is tuned to infinity, is of special interest in various geometries including harmonic \cite{Kinast2005, Nascimbene2010, Ku2012, Sidorenkov2013, Bardon2014} and box traps \cite{Mukherjee2017, Patel2020, Li2022, Yan2024}. It is not only strongly correlated but also an example of scale-invariant quantum many-body system. One of the basic tools used to explore the properties of unitary Fermi gas is the expansion dynamics \cite{Menotti2002} and much insight has been obtained about the role of interactions \cite{Thomas2005, Elliott2014, Deng2016}.

The strongly interacting Fermi gas at finite temperature is described by a hydrodynamic theory (see Eq. \ref{eq:hydrodynamic}), where the transport behaviors are determined by viscosities \cite{Cao2011a, Levin2011viscosity, Shafer2017viscosity}. At unitarity, the bulk viscosity $\zeta_B$ vanishes, and the friction force arise from shear viscosity $\eta$. For a unitary Fermi gas in an anisotropic trap studied previously, the conformal symmetry is broken and the shear viscosity plays an dominant role, which allowed its extraction from expansion dynamics \cite{Cao2011, Thomas2014anomalousViscosity, Thomas2015superfluidViscosity}. On the other hand, for a spherical unitary Fermi gas, the transverse relative motion of the atomic cloud is absent ($\sigma_{ii}=0$), and consequently the effect of the shear viscosity can be neglected. The system without viscosity contribution would have scale invariance of the mean square cloud radius $\left<x^{2}\right> \rightarrow \lambda^{2}\left<x^{2}\right>$ under the transformation $x \rightarrow \lambda x$, being connected to a non-interacting gas. Contrary to the anisotropic system, the spherical unitary Fermi gas has a hidden SO(2,1) symmetry \cite{Werner2006, Supplemental2023} with Hamitonian and raising/lowing operators composing three parts of the SO(2,1) Lie algebra, which predicts the exact relations between trapping potential energy and total energy and the breathing mode with an oscillation frequency twice the trapping frequency. However, the preparation and exploration of the universal properties of a spherical unitary Fermi gas are yet to be demonstrated experimentally.

In this letter, we produce a spherical Fermi gas in an optical dipole trap (ODT) and explore scale invariant behaviors in strongly interacting regimes. By tuning the interaction strength to unitarity, the expansion of the system shows the scale invariance along each direction and at different temperatures, which is absent in an anisotropic system. We find that the trapping potential energy equals to the half of the total energy, verifying the virial theorem at unitarity. Furthermore, we observe the breathing mode with an oscillation frequency twice the trapping frequency and a small damping rate, providing the evidence of SO(2,1) symmetry \cite{Werner2006}. In addition, we explore expansion dynamics away from unitarity when scale invariance is broken, and measure the effective exponent $\gamma$ of $\mu(n)\propto n^\gamma$ where $\mu$ is the chemical potential and $n$ is the average density. To the best of our knowledge, this is the first experiment on the 3D ultracold quantum gases with SO(2, 1) symmetry.

The expansion of strongly interacting Fermi gases is described by the hydrodynamic theory \cite{Cao2011,Elliott2014, Supplemental2023},
\begin{equation} \label{eq:hydrodynamic}
\begin{split}
\frac{d^2}{dt^2}\frac{m{\langle}x_{i}^2{\rangle}}{2}=&{\langle}x_{i}\cdot{\frac{\partial U}{\partial x_{i}}}{\rangle}_{0}+\frac{1}{N}\int{d^3r[\Delta p-\Delta p_0]}\\
&-\frac{1}{N}\int{d^3r(\eta \sigma_{ii}+\zeta_B\sigma')},
\end{split}
\end{equation}
where ${\langle}x_{i}^2{\rangle}$ represents the mean square cloud radius along the $i$th axis ($i=x, y, z$), $U$ is the trapping potential, $t$ is the expansion time, the subscript $(_0)$ denotes the initial condition in the trap at $t=0$, and $\Delta p=p-(2/3)\varepsilon$ is the scale-invariance breaking pressure, where $\varepsilon$ is the energy density \cite{Ho2004}. The last term on the right describes the friction forces arising from shear viscosity $\eta$ and bulk viscosity $\zeta_B$. Here, $\sigma_{ii}=2\dot{b}_{i}/b_{i}-(2/3)\sum_{j}\dot{b}_{j}/b_{j}$ represents the transverse relative motion and $\sigma'=\sum_{i}\dot{b}_{i}/b_{i}$ for the dilation process, where $b_{i}$ denotes the expansion scale factor. In the unitary regime, both $\Delta p$ and $\zeta_B$ vanish \cite{Cao2011, Son2007, Escobedo2009, Dusling2013}. The value of $\sigma_{ii}$ depends on the geometry or symmetry of the atomic cloud. Only for a spherical gas, the relative motion is absent with $\sigma_{ii}=0$, and in this case, we obtain the expansion behavior,
\begin{equation} \label{eq:expansion}
\begin{split}
{\langle}x_{i}^2{\rangle}&={\langle}x_{i}^2{\rangle}_{0}+\frac{t^2}{m}{\langle}x_{i}\cdot{\frac{\partial U}{\partial x_{i}}}{\rangle}_{0}.
\end{split}
\end{equation}
Eq. (\ref{eq:expansion}) shows the ballistic expansion analogous to a non-interacting ideal gas, and the interaction is included in the in-situ atomic cloud size ${\langle}x_{i}^2{\rangle}_{0}$.

The scale-invariant expansion along each direction can be tested by determining the value,
\begin{equation} \label{eq:unitaryvariable}
\begin{split}
\tau_{i}^2(t)=\frac{{\langle}x_{i}^2{\rangle}_t-{\langle}x_{i}^2{\rangle}_{0}}{{\langle}x_{i}^2{\rangle}_{0}\omega^2},
\end{split}
\end{equation}
\noindent where $\omega_{x}=\omega_{y}=\omega_{z}=\omega$ is the trapping frequency. According to Eq. (\ref{eq:expansion}), $\tau^2(t)=t^2$. The ideas of using ballistic expansion and $\tau^2(t)$ as measure of scale invariance were suggested and demonstrated experimentally in an anisotropic Fermi gas \cite{Elliott2014}, where only sum of mean square radii along three axes shows the scale invariance.

\begin{figure}[htbp]
\centerline{\includegraphics[width=8cm]{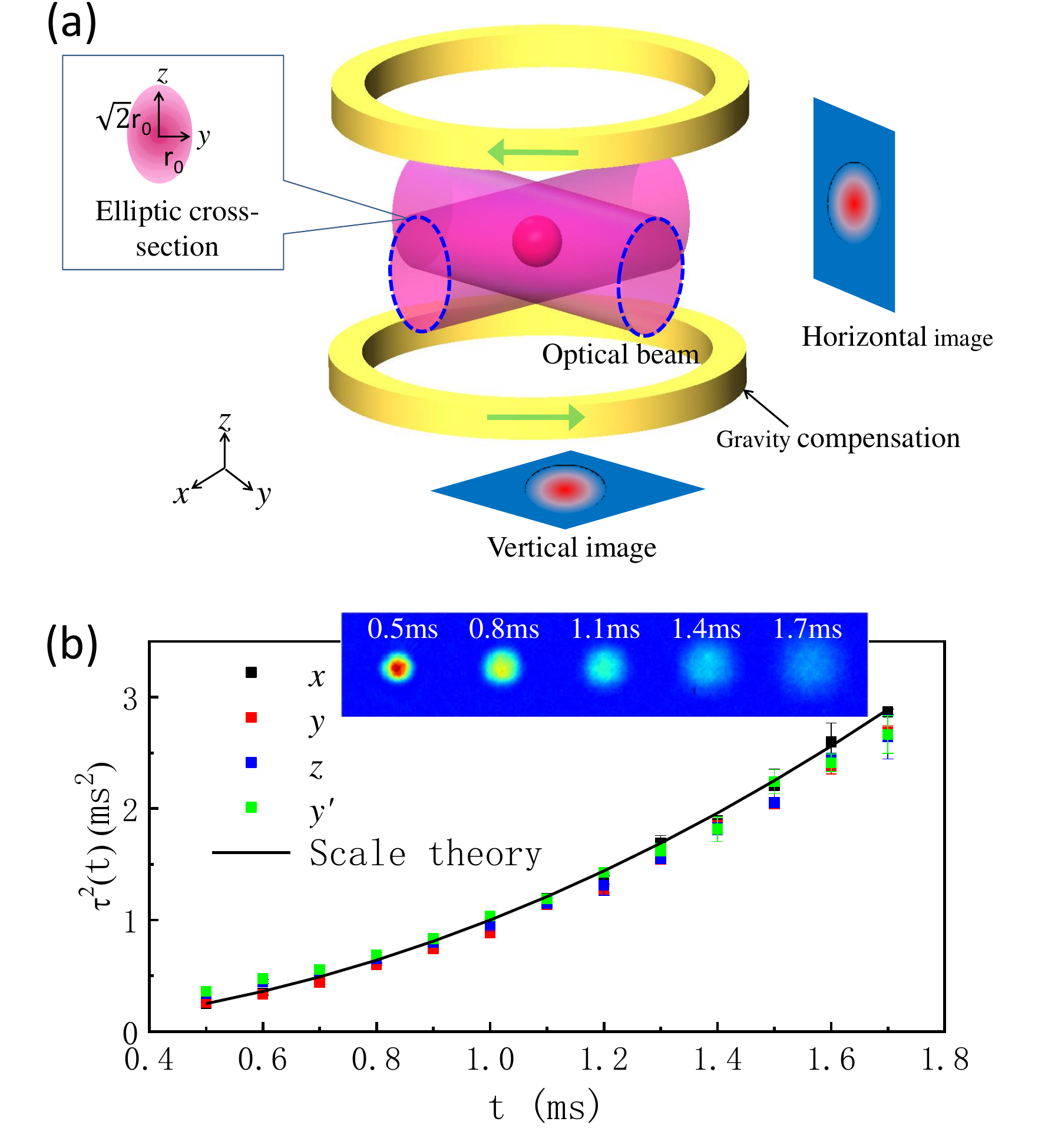}}
\caption{(Colour online). Production of a spherical Fermi gas and scale-invariant expansion along each direction. (a) Schematics of the experimental setup. Two elliptic optical beams, propagating along $x$ and $y$ axis, respectively, form the isotropic trap. The ratio of the beam width along $z$ direction to that in the horizontal plane is $\sqrt{2}$. A pair of anti-Helmholtz coils produce a gradient magnetic field along vertical direction to compensate the gravity. Another pair of Helmholtz coils (not shown) produce a homogeneous magnetic field along vertical direction to tune the interactions. (b) Values of $\tau_i^2(t)$ versus the expansion time $t$ along different directions ($i=x, y, z, y'$). The inset shows typical atomic images taken in the vertical direction, where the expansion time is 0.5 ms, 0.8 ms, 1.1 ms, 1.4 ms and 1.7 ms. $y'$ direction is between $x$ and $y$ axes. The error bar is the standard deviation of several measurements. The black solid curve represents the scale theory $\tau^2(t)=t^2$. $T/T_{\textrm{F}}=$ $0.36(3)$ where $T_{\textrm{F}}$ is the Fermi temperature of the non-interacting Fermi gas.}  \label{Fig1}
\end{figure}

We prepare a spherical Fermi gas based on our previous works \cite{Yan2021, Yan2022}. We initially prepare a $^6$Li atomic degenerate Fermi gas with two spin states $|F=1/2,m_F=\pm1/2\rangle$ in an elongated ODT and at the Feshbach-resonance magnetic field of $834$ G. The experimental setup of the isotropic trap is schematically displayed in Fig. \ref{Fig1}(a), where some special techniques are applied. Firstly, a magnetic field with a gradient $B'_{z}=1.05$ G/cm is applied along $z$ axis to simultaneously compensate the gravity force of the two spin states. This is valid for $^6$Li atoms because the hyperfine interaction is much smaller than the Zeeman shift at the applied magnetic field. Secondly, two elliptic optical beams with a cross-sectional aspect ratio of $\sqrt{2}$, propagating perpendicularly in the horizontal plane, form the isotropic trap. Under these conditions, the trapping frequency can be varied by adjusting the optical power (See Supplemental Material \cite{Supplemental2023} for details). We transfer the unitary Fermi gas to the isotropic trap with an efficiency of more than ${90\%}$. After performing the evaporative cooling in the isotropic trap, we slowly increase the optical power to $3.8$ W in about 75 ms. The temperature is adjusted by controlling the optical power of the evaporative cooling. The atom number is $N=2.9(3)\times10^4$ and the spin polarization is less than 6\%. The trapping frequencies are $(\omega_x, \omega_y, \omega_z)=2\pi\times(1234(6), 1165(11), 1204(3))$ Hz, which are nearly the same along three axes \cite{Supplemental2023}.

We switch off the isotropic ODT and measure the cloud width versus the expansion time $t$ at the magnetic field $B=834$ G. Two laser beams with a frequency difference of 76 MHz, propagating along vertical and horizonal directions, respectively, are applied to detect two spin states. Typical atomic images during the expansion are shown in the inset of Fig. \ref{Fig1}(b), indicating an isotropic expansion in direct contradiction to an elongated Fermi gas \cite{OHara2002}. We use a fringe-removal algorithm \cite{Supplemental2023, Ockeloen2010} to reduce the imaging noise. The cloud radius ${\langle}x_{i}^2{\rangle}_{t}$ is obtained by fitting a Gaussian distribution to the atomic density profile. In the unitary regime, the temperature is determined by analyzing the atomic density distribution, and the cloud radius in the trap ${\langle}x_{i}^2{\rangle}_{0}$ can be theoretically calculated \cite{Kinast2005, Yan2021, Supplemental2023}. Values of $\tau^2(t)$ are calculated according to Eq. (\ref{eq:unitaryvariable}). As shown in Fig. \ref{Fig1}(b), the expansion behaviors along different directions all obey the scale theory $\tau^2(t)=t^2$, which indicates the absence of the effect of viscosity. This scale-invariant expansion along each direction is unique for a spherical Fermi gas.

\begin{figure}[htbp]
\centerline{\includegraphics[width=8.5cm]{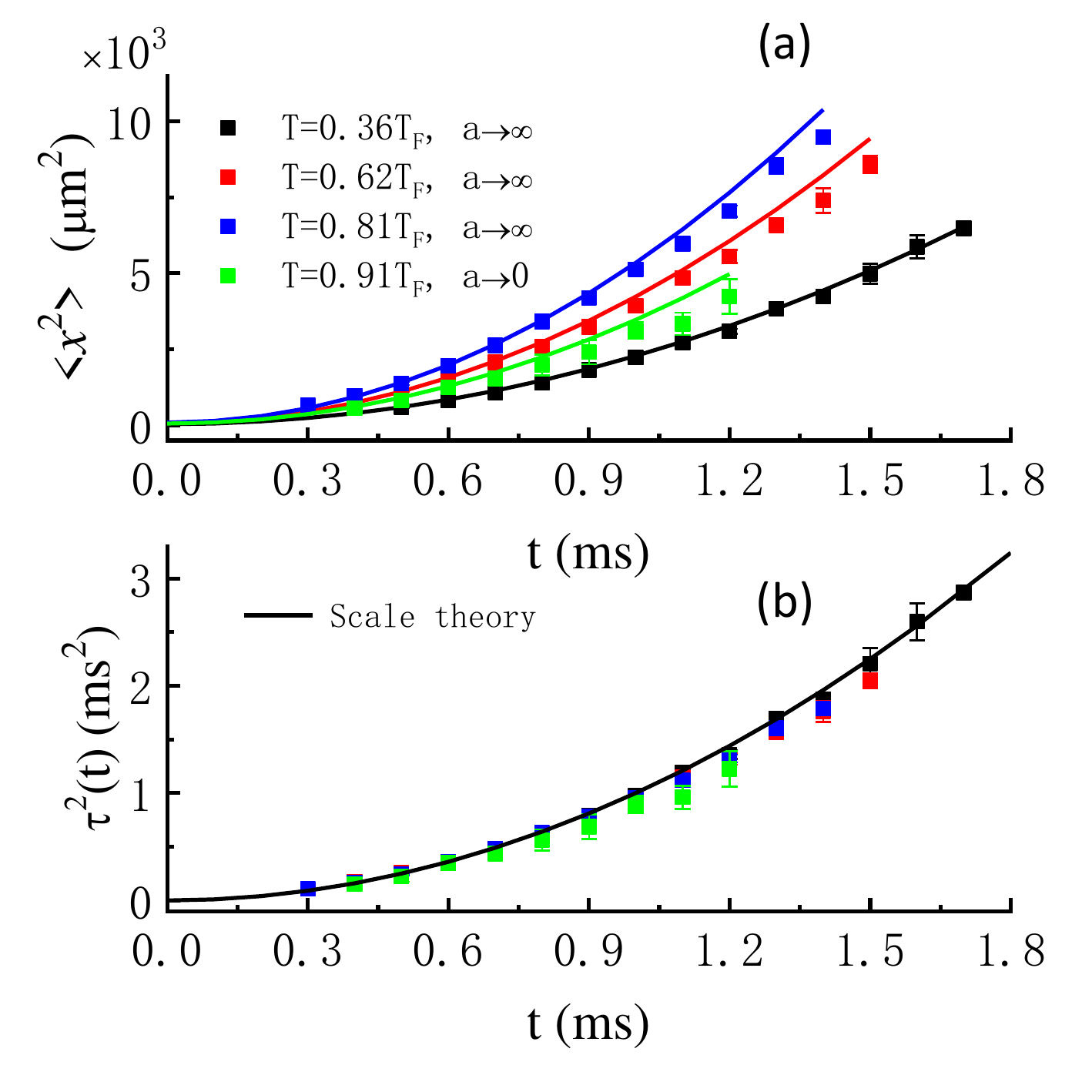}}
\caption{(Colour online). Scale-invariant expansion at different temperatures. (a) Atomic cloud size ${\langle}x^2(t){\rangle}$ versus the expansion time $t$ in the unitary regime ($a \rightarrow \infty$). $T/T_{\textrm{F}}=0.36(3)$ (black), $0.62(2)$ (red) and $0.81(2)$ (blue), respectively. For comparison, the non-interacting Fermi gas ($a\rightarrow0$) is also shown with $T/T_{\textrm{F}}=$ $0.91(3)$ (green). The solid curves represent the calculations of Eq. (\ref{eq:expansion}). The error bar is the standard deviation of several measurements. (b) Values of $\tau^2(t)$ versus the expansion time $t$ at different temperatures. The solid black curve denotes the scale theory $\tau^2(t)=t^2$. } \label{Fig2}
\end{figure}

In Fig. \ref{Fig2}, we measure the atomic expansion at different temperatures. Only expansion along the $x$-axis is displayed for simplicity. Due to the finite-temperature effect, the atomic cloud size ${\langle}x^2(t){\rangle}$ shows an obvious difference, as shown in Fig. \ref{Fig2}(a). System at a higher temperature has a larger in-situ cloud radius ${\langle}x_{i}^2{\rangle}_{0}$, leading to a faster expansion, which agrees well with the theoretical prediction of Eq. (\ref{eq:expansion}). While values of $\tau^2(t)$ at different temperatures are consistent, all obeying the scale theory $\tau^2(t)=t^2$ (see Fig. \ref{Fig2}(b)). For comparison, the expansion behavior of a non-interacting Fermi gas ($a\rightarrow 0$, where $a$ is the $s$-wave scattering length) is also shown. The Fermi gas in the unitary regime ($a\rightarrow\infty$) has the same scaled expansion behavior with that of the non-interacting Fermi gas.

\begin{figure}[htbp]
\centerline{\includegraphics[width=8.0cm]{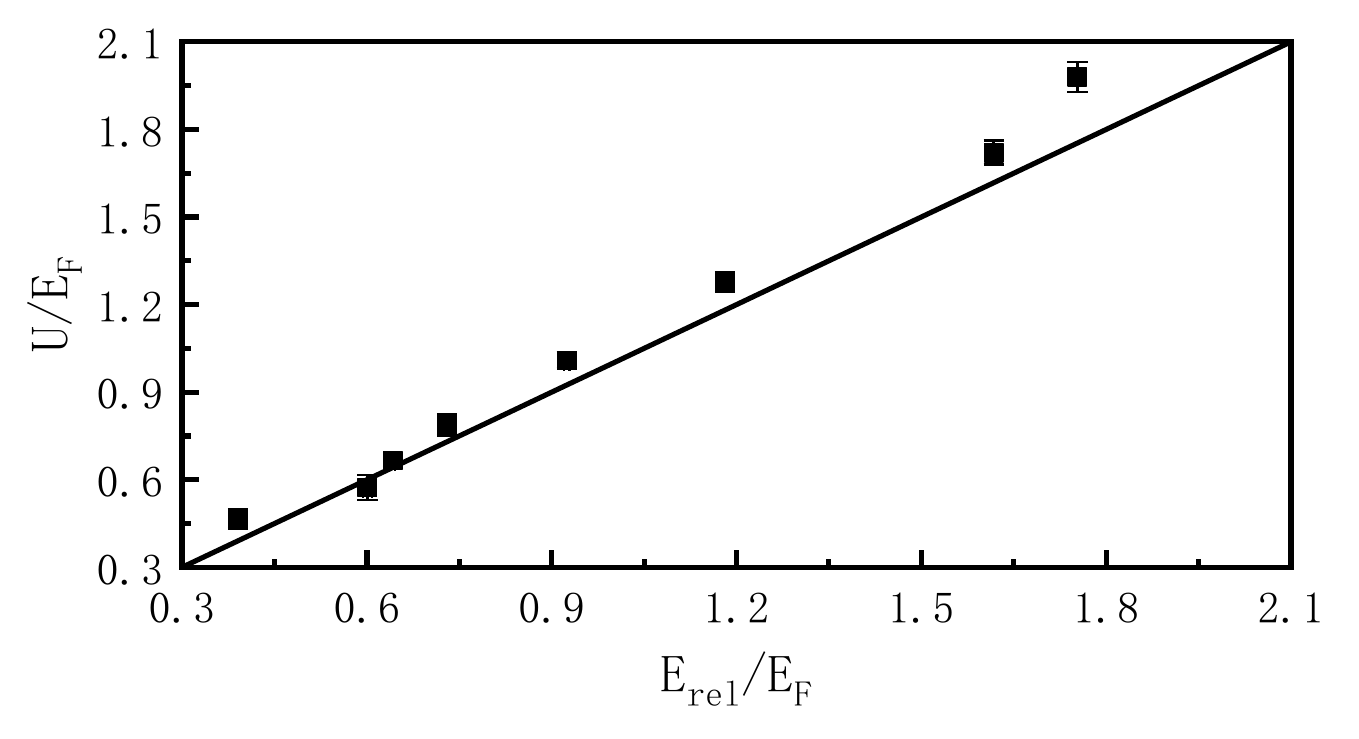}}
\caption{Verifying the virial theorem for a unitary Fermi gas. The trapping potential $U$ shows a linear dependence on the release energy $E_{\textrm{rel}}$. The error bar of the trapping potential comes from the uncertainty of the trapping frequency. The error bar of the release energy from the standard deviation of several measurements is smaller than the mask size. Both energies are normalized by the Fermi energy $E_{\textrm{F}}=(3N)^{1/3}\hbar\omega$, where $N$ is the atom number and $\omega$ is the trapping frequency. The solid line represents the relation $U=E_{\textrm{rel}}$.}  \label{Fig3}
\end{figure}

As predicted by SO(2,1) symmetry, the total energy should be twice the trapping potential energy \cite{Werner2006}. This energy relation, which is called the virial theorem, also could be derived based on the equation of state and verified by measuring the trapping potential energy \cite{Thomas2005}. Here we verify the virial theorem using the expansion method. The total energy of the trapped gas is the sum of trapping potential energy $U$, kinetic energy $E_{\textrm{kin}}$ and interaction energy $E_{\textrm{int}}$, $E_{\textrm{tot}}=U+E_{\textrm{kin}}+E_{\textrm{int}}$. After switching off the trapping potential ($U = 0$), the release energy  $E_{\textrm{rel}}=E_{\textrm{kin}}+E_{\textrm{int}}$ remains constant during the expansion \cite{Stringari1999RMP, Cooper1997PRL, liExpansionDynamicsSpherical2019} and will be completely converted to the kinetic energy in the long-time expansion. So we only need to demonstrate the relation $U=E_{\textrm{rel}}$. By fitting the slope of atomic cloud radius respect to the expansion time, we obtain the release energy $E_{\textrm{rel}}=(3/2)mv_{x}^2$, where $v_{x}$ is the expansion velocity along $x$ axis. We can also determine the trapping potential energy $U=(3/2)m\omega^2 {\langle}x^2{\rangle}_0$, which varies with atomic temperature. The experimental results are shown in Fig. \ref{Fig3}, where the atomic temperature changes across the Fermi degeneracy. The virial theorem is valid over a wide range of energies.

\begin{figure}[htbp]
\centerline{\includegraphics[width=8.0cm]{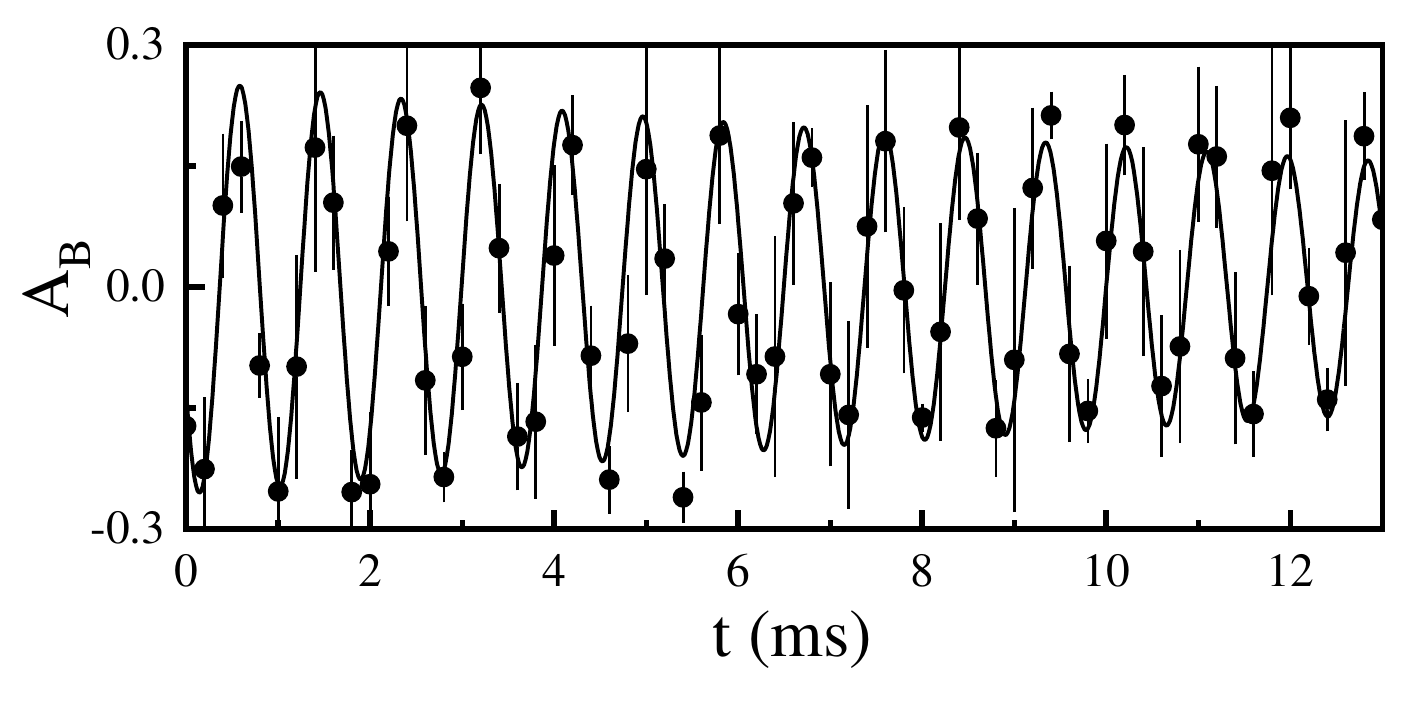}}
\caption{Breathing oscillation as a function of the holding time $t$ in the trap. Experimental data are fitted with a damped sinusoidal function (solid curve). The oscillation frequency is $\omega_B = 2\pi\times1142(2)$ Hz $\approx2\omega_0$ and the damping rate is $\Gamma_B = 38(8)$ $s^{-1}$. Error bar is the standard deviation of several measurements. Here the temperature is $T /T_{F} = 0.29(1)$ and the mean trapping frequency is $\omega_0 = 2\pi\times583$ Hz.}  \label{Fig4}
\end{figure}

Next we will study the breathing mode, demonstrating SO(2,1) symmetry of the system at unitarity. To excite the breathing mode, we sinusoidally modulate the optical field at twice the trapping frequency for 4 periods, making the atomic cloud sizes oscillate in phase along three axes. After different holding times $t$ in the trap, the atomic cloud is imaged with a time of flight of 1 ms. The breathing mode oscillation is defined as
\begin{equation} \label{eq:AB}
A_B=\frac{\sum_i\langle x_i^2\rangle(t)}{\langle\sum_i\langle x_i^2\rangle(t)\rangle}-1,(i=x,y,z),
\end{equation}

\noindent where $\langle x_i^2\rangle(t)$ is the mean square cloud radius and $\langle\sum_i\langle x_i^2\rangle(t)\rangle$ is the average value of all data. As shown in Fig. \ref{Fig4}, the oscillation is fitted with a damped sinusoidal function $A_B=\delta A_B \sin(\omega_Bt+\phi) \exp(-\Gamma_Bt)$. Here the mean trapping frequency along three axes is $\omega_0 = 2\pi\times583$ Hz. The oscillation frequency is $\omega_B=2\pi\times1142(2)$ Hz $\approx2\omega_0$ and the damping rate is $\Gamma_B = 38(8)$ $s^{-1}$. The normalized damping rate to the oscillation frequency is very small, i.e., $\Gamma_B/\omega_B \approx 0.005(1)$. Observation of the breathing mode with an oscillation frequency twice the trapping frequency and a small damping rate provides a direct evidence of SO(2,1) symmetry \cite{Werner2006}. The SO(2,1) symmetry in a 2D quantum gas has also been demonstrated in a similar way \cite{PitaevskiiPRA1997Scale, DalibardPRL2002Transverse, VogtPRL2012Scale}.

\begin{figure}[htbp]
\centerline{\includegraphics[width=8.0cm]{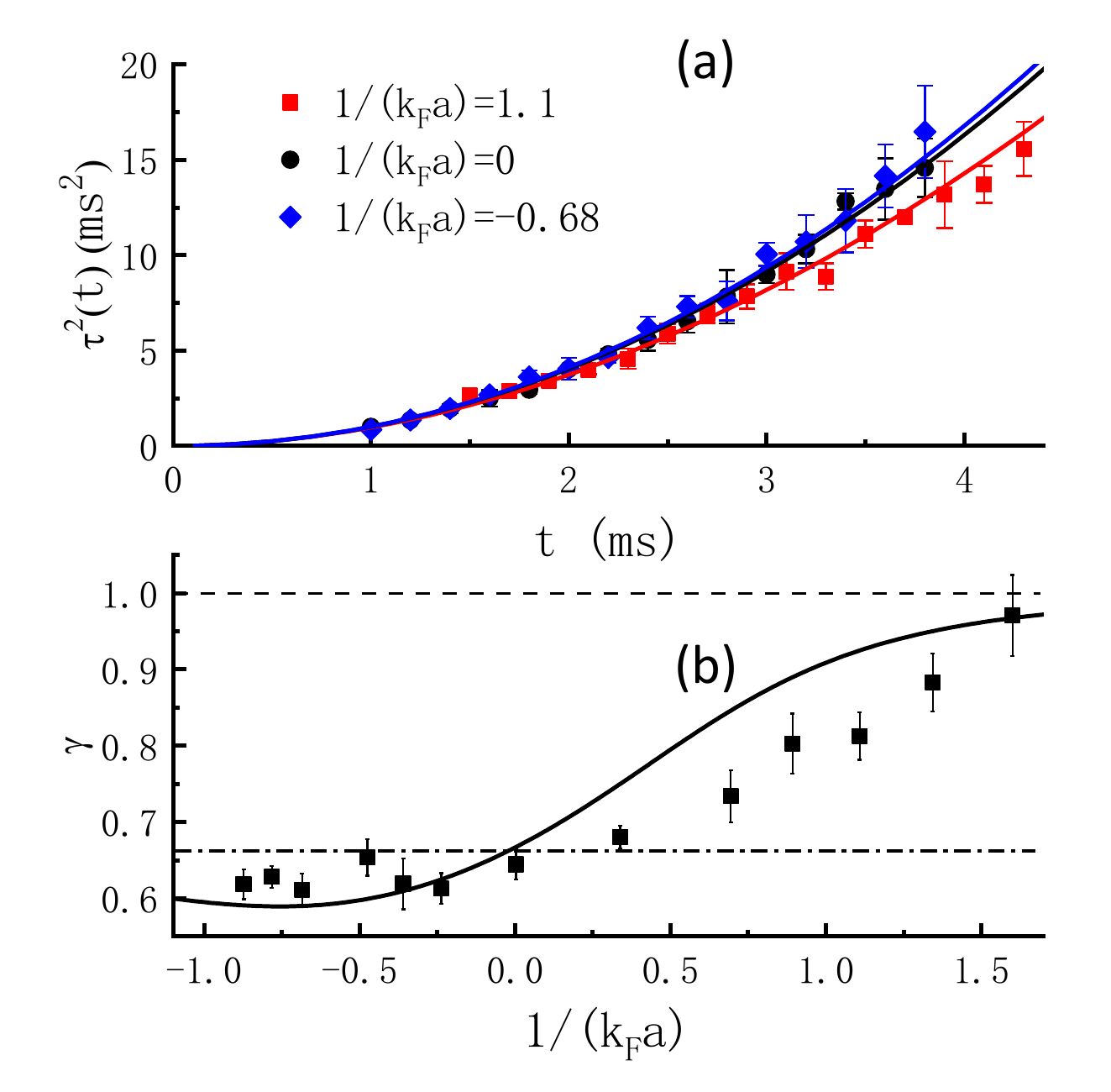}}
\caption{(Colour online). Scale invariance breaking in the BEC-BCS crossover. (a) Values of $\tau^2(t)$ versus the expansion time $t$ at different interactions. The expansion behaviors are represented for BEC ($1/k_{\textrm{F}}a=1.1$, red squares), unitary ($1/k_{\textrm{F}}a=0$, black circles) and BCS ($1/k_{\textrm{F}}a=-0.68$, blue diamonds) regions, respectively. The solid curves denote the calculations of Eq. (\ref{eq:universalvariable}) with experimentally obtained $\gamma$. In the unitary regime, $T/T_{\textrm{F}}=0.21(2)$. (b) Values of $\gamma$ in the BEC-BCS crossover. The solid curve depicts the mean-field calculation at zero temperature as in Ref. \cite{huCollectiveModesBallistic2004}. The dashed line denotes $\gamma=1$ valid deeply in the BEC regime, and the dot-dashed line denotes $\gamma=2/3$ valid at unitarity and deeply in the BCS regime.}  \label{Fig5}
\end{figure}

For comparison, we measure the breathing mode away from the unitarity. On the BEC side ($1/k_{\textrm{F}}a=0.57$), the oscillation frequency is $\omega_B\approx 2.08 \omega_0$ and the normalized damping rate is $\Gamma_B/\omega_B \approx 0.009(1)$. On the BCS side ($1/k_{\textrm{F}}a=-0.84$), the oscillation frequency is $\omega_B\approx 1.94 \omega_0$ and the normalized damping rate is $\Gamma_B/\omega_B \approx 0.006(1)$. Away from the unitarity, the ratio of the oscillation frequency to the trapping frequency is not equal to 2 and the damping rate increases more or less. The difference from the unitarity is large on the BEC side and small on the BCS side, which is similar to the measurements of the free expansion (see Fig. \ref{Fig5}).

Away from the resonance with $\Delta p\neq0$, the scale invariance will be broken. We assume a power-law dependence of the chemical potential, $\mu(n)= n^{\gamma}$, where $n$ is the average atomic density and $\gamma $ is the effective exponent \cite{Stringari2008RMP, huCollectiveModesBallistic2004, Heiselberg2004PRL}. By imposing that the total energy variation vanishes to first order, one gets the energy relation in the BEC-BCS crossover \cite{Stringari2008RMP}, $3\gamma E_{\textrm{rel}}=2U$. Considering that the bulk viscosity is negligibly small \cite{Elliott2014} and the effect of the shear viscosity in a spherical system is zero, we obtain the expansion scale factors \cite{huCollectiveModesBallistic2004},
\begin{equation} \label{eq:scalefactor}
\ddot b_i-(\omega^2/b_i)\Gamma^{-\gamma}=0,
\end{equation}
\noindent where $\Gamma=b^{3}_{i}$. Eq. (\ref{eq:scalefactor}) is a decoupled equation for each direction, where $b_{i}$ can be calculated if knowing the value of $\gamma$. In the unitary regime with $\gamma=2/3$, Eq. (\ref{eq:scalefactor}) has an analytical solution the same as Eq. (\ref{eq:expansion}). In the BEC-BCS crossover, $\tau^2(t)$ can be calculated as
\begin{equation} \label{eq:universalvariable}
\begin{aligned}
\tau^2(t) =\frac{b^2_{i}-1}{\omega^2}.
\end{aligned}
\end{equation}

To measure $\gamma$ at different interactions, we adiabatically ramp the magnetic field from 834 G to the desired value in 300 ms. As the power-law dependence of the chemical potential is valid at zero temperature, we perform the experiment at a low temperature $T/T_{\textrm{F}}=0.21(2)$. The value of $\gamma$ at zero temperature \cite{huCollectiveModesBallistic2004} is initially input into Eq. (\ref{eq:scalefactor}) to calculate $b_{x}(t)$. Then we can obtain the cloud radius in the trap from the expansion data, ${\langle}x^2{\rangle}_0={\langle}x(t)^2{\rangle}_t/b(t)^2_x$, to determine the trapping potential energy $U$. We also measure the release energy $E_{\textrm{rel}}$ from the long-time expansion. Using the energy relation, we obtain a new value of $\gamma$ and input it into Eq. (\ref{eq:scalefactor}) for the next iterative calculation. We repeat the calculation until $|\gamma_{i+1}-\gamma_{i}|/\gamma_{i} \leq10^{-5}$, where $i$ denotes the number of iterations. The obtained $\gamma$ in the BEC-BCS crossover is shown in Fig. \ref{Fig5}(b). In the unitary regime, $\gamma\approx2/3$. On the BEC side, $\gamma$ increases towards the molecule condensate limit with $\gamma=1$. On the BCS side, $\gamma$ decreases to some extent. The experimental measurements have the same variation trend with the mean-field calculation at zero temperature \cite{huCollectiveModesBallistic2004}. The value of $\gamma$ in the unitary regime does not change with temperature. But due to the finite-temperature effect, $\gamma$ is smaller than the zero-temperature calculation on the BEC side, and on the BCS side it is larger. This can be reasonably understood that, as temperature increases, $\gamma$ will change towards the thermal gas with $\gamma=2/3$. $\gamma$ could also be obtained by measuring the equation of state \cite{Navon2010, wangOscillatorylikeExpansionFermionic2020} and collective-mode oscillation \cite{KinastBreakdown2004PRA}.

With obtained $\gamma$, we can determine the cloud radius ${\langle}x^2{\rangle}_0=\gamma E_{\textrm{rel}}/m\omega^2$. Three expansion behaviors in the BEC-BCS crossover are shown in Fig. \ref{Fig5}(a), displaying the obvious deviation from that in the unitary regime ($1/k_{\textrm{F}}a=0$) when the interaction strength is tuned away from the resonance. The expansion is fast on the BCS side ($1/k_{\textrm{F}}a=-0.68$) and slow on the BEC side ($1/k_{\textrm{F}}a=1.1$), which can be well calculated with Eq. (\ref{eq:universalvariable}).

In conclusion, we observe the unique feature of the scale invariance induced by the coexistence of the spherical symmetry and unitary interaction, and demonstrate SO(2,1) symmetry of the system by observing the breathing mode. The virial theorem for the unitary Femi gas has been verified. We also measure the effective exponent $\gamma$ in the equation of state along the BEC-BCS crossover. The spherical unitary Fermi gas provides the platform to study geometrized quantum dynamics with SU(1,1) symmetry \cite{Zhou2020PRLgeometic, Zhou2020PRLbreather}, non-equilibrium dynamics in the presence of conformal symmetry \cite{ZhoufeiPRA2019Conformal, ZhoufeiPRA2020Conformal, ZhoufeiPRL2022Conformal} and the bulk viscosity in the BEC-BCS crossover \cite{Enss2019, Hofmann2020PRAviscosity, Nishida2019AnnalsVisocosity}.

We thank Shizhong Zhang and Xi-Wen Guan for carefully reading and revising the paper, and Georgy Shlyapnikov for favorite discussions. This work has been supported by the National Key R \&D Program under Grant No. 2022YFA1404102, NSFC (Grant Nos. U23A2073, 12374250 and 12121004), CAS under Grant No. YJKYYQ20170025, and Hubei province under Grant No. 2021CFA027.

\newpage
\begin{widetext}

\setcounter{secnumdepth}{3} 

\setcounter{equation}{0}

\setcounter{figure}{0}
	
\renewcommand{\thefigure}{S\arabic{figure}}
\renewcommand{\thetable}{S\arabic{table}}
\renewcommand{\theequation}{S\arabic{equation}}

\section*{Supplemental materials}

\section{Production of an isotropic optical trap}
\subsection{Experimental setup of the isotropic optical trap}

Experimental setup of the isotropic trap is schematically displayed in Fig. \ref{SFig1}, which is  also shown as Fig. 1(a) in the main text. Here we will display more details about the setup. We prepare $^6$Li atomic degenerate Fermi gas with two spin states $|F=1/2,m_F=\pm1/2\rangle$. A pair of Helmholtz coils (not shown in the figure) produce a homogeneous magnetic field in vertical direction, which is used to tune the interactions. The Feshbach-resonance magnetic field for $^6$Li atoms is about $834$ G. Some special techniques are applied to produce the isotropic trap. Firstly, A pair of anti-Helmholtz coils produce a gradient magnetic field in vertical direction to simultaneously compensate the gravity force of the two spin states, which is valid for $^6$Li atoms because the hyperfine interaction is much smaller than the Zeeman shift at the strong magnetic field applied in the experiment. The magnetic gradient is $B'_{z}=1.05$ G/cm. Secondly, two elliptic optical beams with a cross-sectional aspect ratio of $\sqrt{2}$, propagating perpendicularly to each other, form the isotropic trap. To avoid interference, the two ODT (optical dipole trap) beams have orthogonally linear polarizations and a frequency difference of 220 MHz. Under these conditions, the trapping frequency can be varied by the optical power. The waist radius of the optical beam is 60 $\mu$m and the wavelength is 1064 nm. Two laser beams with a frequency difference of 76 MHz, propagating in vertical and horizonal directions, respectively, are applied to detect two spin states separately.

We transfer the unitary Fermi gas to the isotropic trap with an efficiency of more than ${90\%}$, by lowering the power of the elongated ODT and simultaneously increasing that of the isotropic ODT in a period of 25 ms. After performing the evaporative cooling in the isotropic trap, we slowly increase the optical power to $3.8$ W in about 75 ms. The temperature is adjusted by controlling the optical power of the evaporative cooling. The atom number is $N=2.9(3)\times10^4$ and the spin polarization is less than 6\%. The cloud radius ${\langle}x_{i}^2{\rangle}_{t}$ is obtained by fitting a Gaussian distribution to the atomic density profile, which is corrected by the fit to the finite-temperature Thomas-Fermi function at low temperatures. In measuring the expansion behaviors, the final optical power of the trap is $3.8$ W and the trapping frequencies are $(\omega_x, \omega_y, \omega_z)=2\pi\times(1234(6), 1165(11), 1204(3))$ Hz. In measuring the effective exponent $\gamma$, the trapping frequencies are $(\omega_x,\omega_y,\omega_z)=2\pi\times (188(1), 176(2), 187(1))$ Hz. And in measuring the breathing mode, the mean trapping frequency along three axes is $\omega_0=2\pi\times 583$ Hz.

\begin{figure}[htbp]
\centerline{\includegraphics[width=15cm]{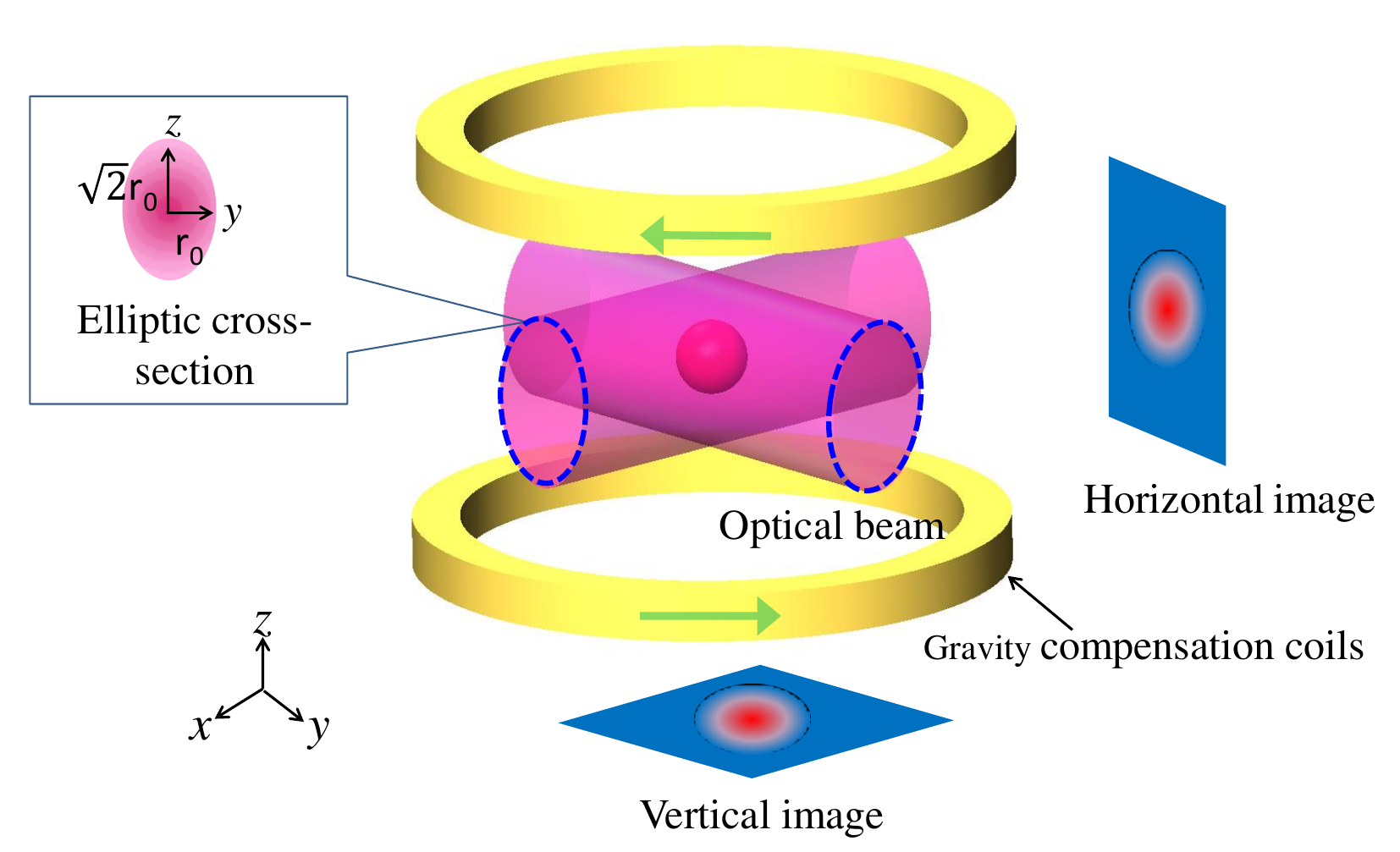}}
\caption{ Experimental setup of the isotropic  trap. Two elliptic optical beams with a cross-sectional aspect ratio of $\sqrt{2}$, propagating perpendicularly to each other, form the isotropic trap. A pair of anti-Helmholtz coils produce a gradient magnetic field in vertical direction to compensate the gravity. Another pair of Helmholtz coils (not shown) produce a homogeneous magnetic field in vertical direction to tune the interactions. Two laser beams with a frequency difference of 76 MHz, propagating in vertical and horizonal directions, respectively, are applied to detect two spin states separately.}   \label{SFig1}
\end{figure}

\subsection{Production of the elliptic beam with a cross-sectional aspect ratio of $\sqrt{2}$}

To form an isotropic optical trap, we should produce two elliptical beams whose cross-sectional aspect ratio is $\sqrt{2}$. The optical configuration to produce one beam is shown in Fig. \ref{SFig2}. A 1064 nm laser beam outputs from a polarization-maintaining fiber. The Glan-Taylor prism is used to purify the optical polarization. A set of three cylindrical lens increases the radius in $x$ direction, while maintains that in $z$ direction. The focus lengths and distances of the lens are carefully selected to obtain the desired aspect ratio of the cross-section, $w_{z}:w_{x}=1:\sqrt{2}$. After passing the final achromatic doublets, the aspect ratio reverses, i.e., $w_{z}:w_{x}=\sqrt{2}:1$. In order to increase the optical stability, the optical devices are constructed with stainless steel mounts, and the experimental setup rests on an air-floating platform.

\begin{figure}[htbp]
\centerline{\includegraphics[width=17cm]{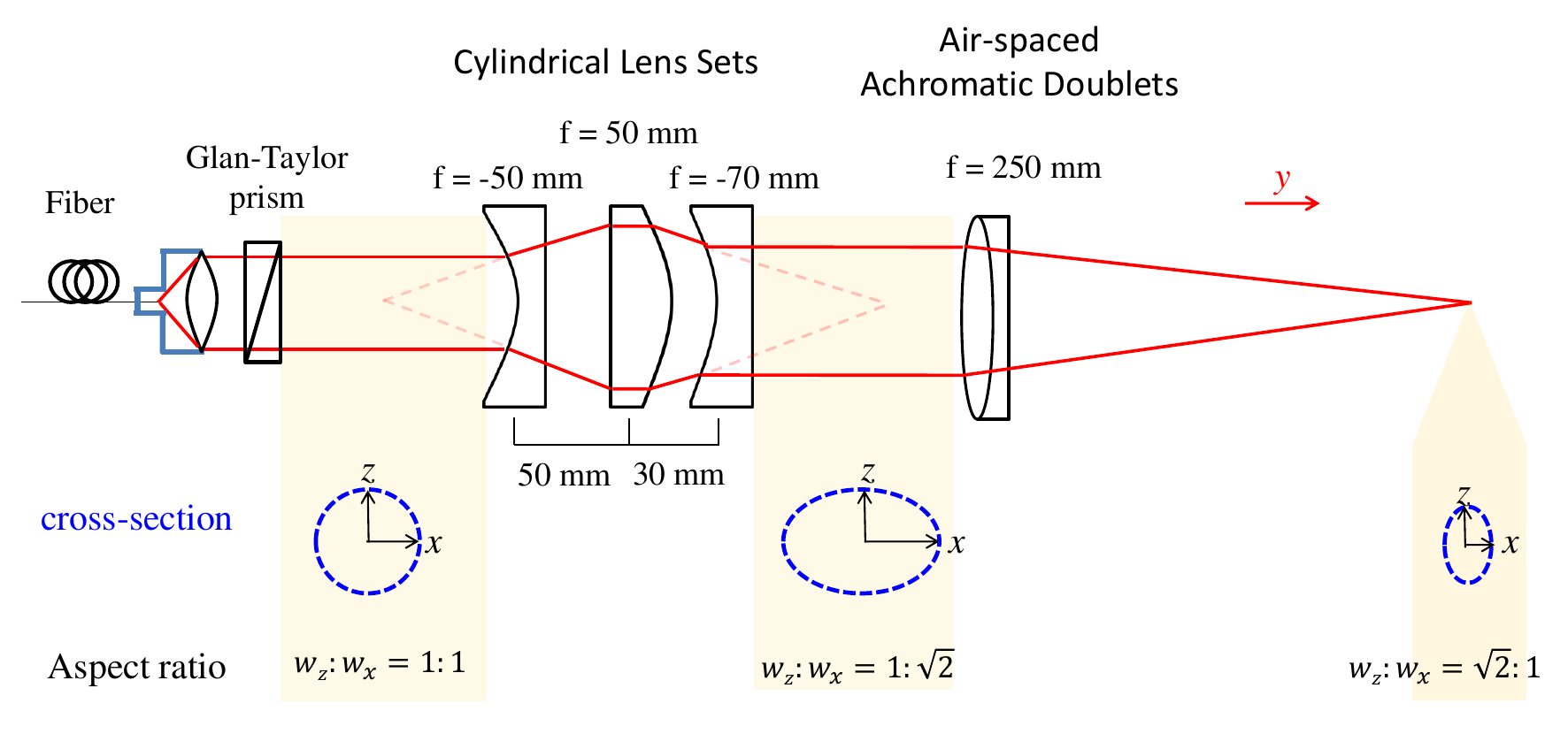}}
\caption{Optical configuration for preparing the elliptical beam. The optical beam propagates in $y$ direction. The first row shows the optical devices, the second row schematically displays the shape of the cross-section on $x-z$ plane, and the third row indicates the value of the cross-sectional aspect ratio $w_{z}:w_{x}$. $w_{x, z}$ is the waist radius in $x$ ($z$) direction. After the output of the optical fiber, the cross-section has a circular shape, i.e., $w_{z}:w_{x}=1:1$. A set of three cylindrical lens increases the radius in $x$ direction, while maintains that in $z$ direction, resulting an elliptical cross-section with $w_{z}:w_{x}=1:\sqrt{2}$. The focus lengths of the lens are -50 mm, 50 mm, and 70 mm, respectively. The distances between lens are 50 mm and 30 mm, respectively. The final achromatic doublets with a focus length of 250 mm reverses the aspect ratio, i.e., $w_{z}:w_{x}=\sqrt{2}:1$.} \label{SFig2}
\end{figure}

\subsection{Gravity compensation with a gradient magnetic field}

To study interactions between two spin states in an isotropic optical trap with variable trapping frequencies, the gravity force of the two spin states should be simultaneously compensated. For $^6$Li atoms at the Feshbach resonant magnetic field of $834$ G, the hyperfine interaction ($\approx 228$ MHz) is much smaller than the Zeeman shift ($\approx 1.2$ GHz). In this condition, the hyperfine quantum number F is no longer a good quantum number. The energy shift due to the Zeeman effect at this strong magnetic field is

\begin{equation} \label{20}
\begin{split}
\Delta E=\mu_{\textrm{B}}(g_J m_J+g_I m_I)B_z,
\end{split}
\end{equation}
\noindent where $g$ is the $Land\acute{e}$ factor and $\mu_\textrm{B}$ is the Bohr magneton. For the two lowest spin states $|F=1/2,m_F=\pm1/2\rangle$, $L=0$ and $m_{J}=m_{S}=-1/2$. With $g_I\ll g_S$, the nuclear contribution can be neglected. Then the energy shift is well approximated by
\begin{equation} \label{21}
\begin{split}
\Delta E\simeq\mu_\textrm{B} g_S m_S B_z,
\end{split}
\end{equation}
\noindent where $g_S=2$. The atoms in two spin states have the same magnetic moment $\mu=m_S g_S\mu_\textrm{B}=-\mu_\textrm{B}$. To compensate the gravity force of the two spin states, the magnetic field gradient $B'_{z}$ can be calculated,
\begin{equation} \label{eq:gravitycompensation}
\begin{split}
\mu_\textrm{B} B_z'=mg.
\end{split}
\end{equation}

In the experiment, we use a pair of anti-Helmholtz coils in $z$ direction to generate a quadrupole magnetic field, which is combined with the Feshbach magnetic field to create a linear magnetic field in $z$ direction. According to Eq. (\ref{eq:gravitycompensation}), $B_z'=1.05$ G/cm.

\subsection{Residual Confinement of the Feshbach Magnetic Field}

We probe the atomic expansion at a magnetic field of 834 G. So it is important to analyze the residual confinement of the Feshbach magnetic field. A pair of Helmholtz coils in $z$ direction are used to tune the Feshbach resonance. The Feshbach magnetic field has a curvature which gives rise to an additional trapping potential,
\begin{equation} \label{23}
\begin{split}
U_{\textrm{mag}}=\frac{1}{2}\mu_\textrm{B} B_z''z^2.
\end{split}
\end{equation}

\noindent Then the trap frequency due to the magnetic field can be calculated,
\begin{equation} \label{24}
\begin{split}
\omega_{\textrm{mag}}=\sqrt{\frac{\mu_\textrm{B} B_z''}{m}}.
\end{split}
\end{equation}

Through the parameters of the Feshbach coils, the curvature of the magnetic field in the vertical direction is calculated to be $B''_{z}=0.1\ \textrm{G/cm}^2$ , while on the horizonal plane, it is negligibly small, i.e., $B''_{\perp}\rightarrow0\ \textrm{G/cm}^2$. Then the residual trapping frequency originating from the Feshbach magnetic field is only $\omega_{\textrm{mag}}=2\pi\times1.54$ Hz, which is much smaller than that of the optical dipole trap. According to Ref. \cite{ZhoufeiPRA2020Conformal}, the scale factor $b(t)$ of the atomic cloud after released from the optical trap in z-direction is
\begin{equation} \label{25}
\begin{split}
b(t)=\left[\cos^2(\omega_{\textrm{mag}}t)+\frac{\omega^2_{\textrm{opt}}}{\omega^2_{\textrm{mag}}}\sin^2(\omega_{\textrm{mag}}t)\right]^{1/2}
\end{split}
\end{equation}
where $\omega_{\textrm{opt}}$ is the trapping frequency of the optical trap and $t$ is the expansion time in the magnetic field. When $\omega_{\textrm{mag}}= 0$, the expansion evolves according to $b_0(t)=(1+\omega^2_{\textrm{opt}}t^2)^{1/2}$, which corresponds to the scale invariant process. The effect of the residual magnetic confinement can be defined by $\delta_{b}=(b_0(t)-b(t))/b_0(t)$, which represents the shift of the atomic cloud size.

In the experiment, the trapping frequency of the optical trap is $\omega_{\textrm{opt}}\approx2\pi\times1201$ Hz. For $t=2$ ms, the longest expansion time in the experiment, $\delta_{b}\sim 6\times10^{-5}$. So the residual confinement effect of the Feshbach magnetic field can be ignored.

\subsection{Theoretical analysis on how to form an isotropic optical dipole trap.}
\subsubsection{One Gaussian Beam}
We first consider one focused Gaussian beam. Suppose that the optical beam propagates along $z$ axis. Then the optical intensity is
\begin{equation} \label{1}
\begin{split}
I(r,z)=\frac{I_0}{1+(z/z_0)^2}\exp\left[\frac{-2r^2}{w_0^2}\right],
\end{split}
\end{equation}
where $I_0$ is the peak intensity, $z_0=\pi w_0^2/\lambda$ is the Rayleigh length, and $w_0$ is waist radius (the $1/e^2$ intensity radius of the beam at the focus). Then the trapping potential is given by
\begin{equation} \label{eq:onebeamtrap}
\begin{split}
U(r,z)=-\frac{U_0}{1+(z/z_0)^2}\exp\left[\frac{-2r^2}{w_0^2}\right],
\end{split}
\end{equation}

\noindent where
\begin{equation} \label{3}
\begin{split}
&U_0=\frac{3\pi c^2\Gamma}{2\omega_0^3\Delta\omega}I_0 ,\\
&I_0=\frac{2P}{\pi w_0^2}.
\end{split}
\end{equation}

\noindent Here $c$ is the speed of light, $\omega_0=2\pi c/\lambda_0$, $\omega=2\pi c/\lambda$, and $\Delta\omega=\omega_0-\omega$. In the experiment, the natural line width is $\Gamma$=$2\pi\times$ 5.87 MHz for $^6$Li atom, $\lambda_0=671$ nm and $\lambda=1064$ nm. To determine the trapping frequency of the trap, we expand the trapping potential of Eq. (\ref{eq:onebeamtrap}) into Taylor series around the center $(x, y, z)=(0, 0, 0)$,
\begin{equation} \label{4}
\begin{split}
U(r,z)\simeq-U_0(1-\frac{z^2}{z_0^2})(1-\frac{2r^2}{w_0^2})+\ldots
\\
\simeq-U_0+\frac{U_0}{z_0^2}z^2+2\frac{U_0}{w_0^2}r^2+\ldots.
\end{split}
\end{equation}

\noindent Then
\begin{equation} \label{5}
\begin{split}
&\frac{U_0}{z_0^2}z^2=\frac{1}{2}m \omega_{\textrm{axial}}^2z^2,\\
&2\frac{U_0}{w_0^2}r^2=\frac{1}{2}m \omega_{\textrm{radial}}^2r^2,
\end{split}
\end{equation}
\noindent where $\omega_{\textrm{axial}}$ and $\omega_{\textrm{radial}}$ are the trapping frequencies of the optical beam in axial and radial directions, respectively.
\begin{equation} \label{6}
\begin{split}
&\omega_{\textrm{axial}}=\sqrt{\frac{2U_0}{m z_o^2}},\\
&\omega_{\textrm{radial}}=\sqrt{\frac{4U_0}{m w_o^2}}.
\end{split}
\end{equation}

In the experiment, the waist radius $w_o$ is 60 $\mu$m. Then $\omega_{\textrm{axial}}/\omega_{\textrm{radial}}\simeq 4.0\times10^{-3}$. The trapping effect in axial direction of the optical beam is negligibly small.

\subsubsection{Two Orthogonal Beams with the Circular Cross-Section}

We consider two identical Gaussian beams with the circular cross-section. The two optical beams propagate along $x$ and $y$ axes, respectively. Then the trapping potential can be expressed as
\begin{equation} \label{eq:twobeamtrap1}
\begin{split}
U_{\textrm{OPT}}= U_{\textrm{OPT1}}+U_{\textrm{OPT2}}
=-\frac{U_0}{1+(x/x_0)^2}\exp\left[-\frac{2(y^2+z^2)}{w_0^2}\right]-
\frac{U_0}{1+(y/y_0)^2}\exp\left[-\frac{2(x^2+z^2)}{w_0^2}\right],
\end{split}
\end{equation}
\noindent where $x_0=y_0$ is the Rayleigh length and $ w_0$ is the waist radius.

We expansion Eq. (\ref{eq:twobeamtrap1}) into Tailor series at the point $(x, y, z)=(0, 0, 0)$,

\begin{equation} \label{8}
\begin{split}
U_{OPT}\simeq-2U_0+\left(\frac{U_0}{x_0^2}+2\frac{U_0}{w_0^2}\right)x^2+\left(\frac{U_0}{y_0^2}+2\frac{U_0}{w_0^2}\right)y^2+4\frac{U_0}{w_0^2}z^2+\ldots.
\end{split}
\end{equation}

As mentioned above, the trapping effect in the axial direction of the optical beam can be ignored, and we only consider the trapping effect in the radial direction. Then the trapping frequencies can be calculated as

\begin{equation} \label{eq:twobeamtrapfrequency1}
\begin{split}
\omega_x=\omega_y\simeq\omega_{\textrm{radial}},\\
\omega_z\simeq\sqrt{2}\omega_{\textrm{radial}}.
\end{split}
\end{equation}
According to Eq. (\ref{eq:twobeamtrapfrequency1}), the trapping frequency in the vertical direction is about $\sqrt{2}$ times that in the horizontal direction.

\subsubsection{Two Orthogonal Beams with the Elliptic Cross-Section}

According to the above analysis, we couldn't form an isotropic trap using the beams with the circular cross-section. Here we consider two beams with the elliptical cross-section. The two optical beams propagate along $x$ and $y$ axes, respectively. The waist radius in $x$ and $y$ direction is different from that in $z$ direction, i.e., $w_{x0}=w_{y0}=w_0$, $w_{x0} \neq w_{z0}$. Ignoring the trapping effect in the axial direction of the optical beam as mentioned above, then the trapping potential is calculated as
\begin{equation} \label{eq:twobeamtrap2}
\begin{split}
U_{\textrm{OPT}}=&U_{\textrm{OPT1}}+U_{\textrm{OPT2}}
\approx -U_0\exp\left[-2\left(\frac{y^2}{w_0^2}+\frac{z^2}{w_{z0}^2}\right)\right]- U_0\exp\left[-2\left(\frac{x^2}{ w_0^2}+\frac{z^2}{ w_{z0}^2}\right)\right].
\end{split}
\end{equation}

Ignoring the trapping effect in the axial direction of the optical beam, we expand Eq. (\ref{eq:twobeamtrap2}) into Tailor series at the point $(x, y, z)=(0, 0, 0)$,
\begin{equation} \label{12}
\begin{split}
U_{\textrm{OPT}}\simeq-2U_0+2\frac{U_0}{w_0^2}x^2+2\frac{U_0}{ w_0^2}y^2+4\frac{U_0}{w_{z0}^2}z^2+\ldots.
\end{split}
\end{equation}

Then the trapping frequencies can be calculated,
\begin{equation} \label{eq:twobeamtrapfrequency2}
\begin{split}
\omega_x=\omega_y \simeq \omega_{\textrm{radial}},\\
\frac{\omega_z}{\omega_x}=\frac{\sqrt{2}w_0}{ w_{z0}}.
\end{split}
\end{equation}

\noindent According to Eq. (\ref{eq:twobeamtrapfrequency2}), if $w_{z0}=\sqrt{2} w_0$, for any optical power, the trapping frequencies along three orthogonal axes are the same,
\begin{equation} \label{eq:threetrapfrequency}
\begin{split}
\omega_x=\omega_y = \omega_z.
\end{split}
\end{equation}

In conclusion, in our experiment, the gravity force is compensated with a gradient magnetic field, and the residual confinement of the Feshbach magnetic field is negligibly small. Then we can form an isotropic optical trap using two orthogonal beams with the elliptical cross-section. The aspect ratio of the cross-section is $\sqrt{2}$. The trapping frequency can be varied by the power of the optical beam.

\section{Measurement of the trapping frequency}

In the experiment, the trapping frequency of the trap is determined by measuring the center-of-mass oscillation of the atomic cloud. We perturb the position of the atomic cloud in the spherical trap by controlling a pulse of the elongated optical trap. We tune the relative position of the elongated trap to the spherical trap, simultaneously shifting positions of the atomic cloud in three axes. After switching off the pulse of the elongated trap, the atoms will oscillate in the trap. After different waiting time in the trap, we probe atoms with a time-of-flight (TOF) of 1 ms. The atomic temperature is above the superfluid temperature $T_{\textrm{c}}$, and the atomic position $x_{i}$ ($i \rightarrow x, y, z$) is obtained by fitting the density profile using a Gaussian distribution. We can determine the oscillation of the center-of-mass $\Delta x_{i}(t)=x_{i}(t)-\bar{x}_{i}$, where $t$ is the waiting time in the trap, and $\bar{x}_{i}$ is the mean value of $x_{i}(t)$. Then we can fit $\Delta x_{i}(t)$ using a sinusoidal function,
\begin{equation} \label{eq:dipolemode}
\begin{split}
\Delta x_{i}(t)=A_{i}\sin(\omega_{i}t+\phi_{i}),
\end{split}
\end{equation}
\noindent where $A_{i}$ is the oscillation amplitude, and $\omega_{i}$ is the trapping frequency in $x_{i}$ direction.

Fig. \ref{Sfig3} shows the measurement results, where the optical power is 3.8 W. The trapping frequencies are $(\omega_x, \omega_y, \omega_z)=2\pi\times(1234(6), 1165(11), 1204(3))$ Hz, which are almost the same along three axes.

\begin{figure}[htbp]
\centerline{\includegraphics[width=17cm]{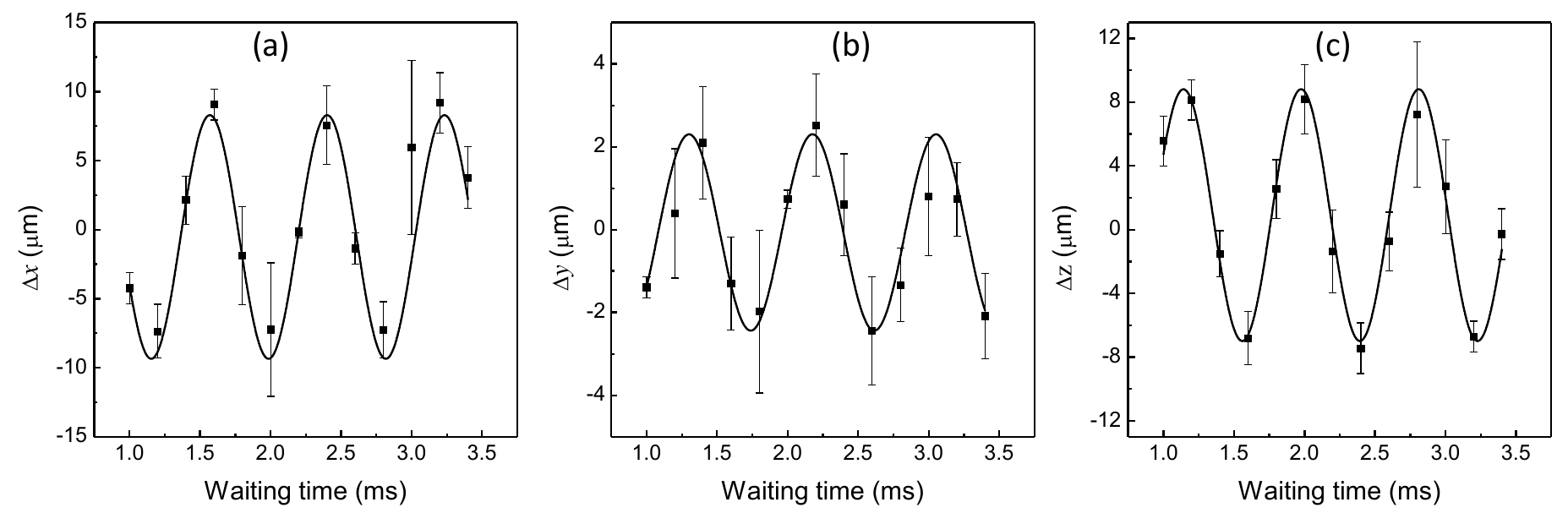}}
\caption{Center-of-mass oscillations of the atomic cloud. $\Delta x_{i}$ ($i \rightarrow x, y, z$) denotes the distance away from the mean value of the atomic position. The error bar is the standard deviation of three measurements. The black solid line denotes the sinusoidal fitting.}  \label{Sfig3}
\end{figure}

\section{Measurement of the temperature and in-situ atomic cloud size}
\subsection{How to obtain the temperature of the unitary Fermi gas}
The temperature of the unitary Fermi gas is obtained using the method similar to J. E. Thomas \cite{Kinast2005} and C. J. Vale's groups \cite{veeravalliBraggSpectroscopyStrongly2008a}. The normalized one-dimensional profile of a non-interacting Fermi gas is

\begin{equation} \label{44}
n(r_{i}(t),T)=-\frac{3N}{\sqrt{\pi}\sigma_{\textrm{F}i}(t)}\left(\frac{T}{T_{\textrm{F}}}\right)^{5/2}
\textrm{Li}_{5/2}\left[-\exp\left(\frac{\frac{\mu}{E_{\textrm{F}}}-\frac{r_{i}^{2}}{\sigma_{\textrm{F}i}^{2}(t)}}{T/T_{\textrm{F}}}\right)\right],
\end{equation}
\noindent where $N$ is the total atom number, $\textrm{Li}_{5/2}$ is the polylogarithm function, $\mu$ is the chemical potential and $\sigma_{\textrm{F}i}(t)$ is the Thomas-Fermi radius of the atomic cloud after released from the trap.

The fitting procedure for the unitary Fermi gas is similar to that of the non-interacting gas as mentioned above, except that the Fermi radius $\sigma_{\textrm{F}i}$ and Fermi temperature $T_{\textrm{F}}$ should be replaced by $\sigma_{\textrm{F}i}^{*}$ and $T_{\textrm{F}}^{*}$ , respectively. At unitarity, $\sigma_{\textrm{F}i}^{*}=(1+\beta)^{1/4}\sigma_{\textrm{F}i}$ and $T_{\textrm{F}}^{*}=(1+\beta)^{1/2}T_{\textrm{F}}$ , where $\beta$ is a universal constant, and the fitting profile becomes

\begin{equation} \label{45}
n(r_{i}(t),T)=-\frac{3N}{\sqrt{\pi}\sigma_{\textrm{F}i}^{*}(t)}\left(\tilde{T}\right)^{5/2}\textrm{Li}_{5/2}\left[-exp\left(q-\frac{r_{i}^{2}(t)}{(\sigma_{\textrm{F}i}^{*}(t))^{2}\tilde{T}}\right)\right],
\end{equation}

\noindent where $q={\mu}/(E_{\textrm{F}}\tilde{T})$ , $E_{\textrm{F}}=\hbar\omega_{0}(3N)^{1/3}$ and $\tilde{T}$ is the empirical temperature given by

\begin{equation} \label{46}
\tilde{T}=\frac{T}{T_{\textrm{F}}\sqrt{1+\beta}}.
\end{equation}

The expansion behavior is known in the unitary regime. $\sigma_{\textrm{F}i}^{*}(t)$ can be calculated by the atom number and trapping frequency. Another way to obtain $\sigma_{\textrm{F}i}^{*}(t)$ is to fit the atomic density profile using the zero-temperature Thomas-Fermi distribution at the lowest temperature. We fix $\sigma_{\textrm{F}i}^{*}(t)$ constant for the fits at all higher temperatures, leaving only $\tilde{T}$ and $q$ as the free parameters. Then we obtain the value of $T/T_{\textrm{F}}$ from Eq. (\ref{46}), where $\beta=-0.56$ \cite{luo2009thermodynamic}.

\subsection{How to obtain the in-situ atomic cloud size of the unitary Fermi gas in finite temperature}

At finite temperature, for a non-interacting Fermi gas, the in-situ mean square size $\left\langle r_{i}^{2}(T)\right\rangle _{0}$ is given by
\begin{equation} \label{47}
\left\langle r_{i}^{2}(T)\right\rangle _{0}=\frac{\sigma_{\textrm{F}i}^{2}}{8}\frac{E}{E_{0}}\left(\frac{T}{T_{F}}\right)
\end{equation}

\noindent where $\sigma_{\textrm{F}i}=\sqrt{{2E_{F}}/{m\omega_{0}^{2}}}$ is the Fermi radius of the atomic cloud in the trap, $E_0$ is the ground energy.

For the unitary Fermi gas,

\begin{equation} \label{48}
\left\langle r_{i}^{2}(\tilde{T})\right\rangle _{0}=\frac{(\sigma_{\textrm{F}i}^{*})^{2}}{8}\frac{E}{E_{0}}(\tilde{T}).
\end{equation}

As seen in Ref. \cite{Kinast2005}, with the same value of $T/T_{F}$ and $\tilde{T}$ for the non-interacting Fermi gas and unitary Fermi gas, $\frac{E}{E_{0}}(\frac{T}{T_{F}})=\frac{E}{E_{0}}(\tilde{T)}$. After knowing the temperature $\tilde{T}$ of the unitary Fermi gas using the method mentioned above, the value of $\frac{E}{E_{0}}(\tilde{T)}$ can be obtained. According to Eq. (\ref{48}) and the relation $\sigma_{\textrm{F}i}^{*}=(1+\beta)^{1/4}\sigma_{\textrm{F}i}$ , we can obtain the in-situ size $\left\langle r_{i}^{2}\right\rangle _{0}$ of the unitary Fermi gas in finite temperature.

\section{Optimization of the Atomic image using the fringe-removal analysis}

To obtain the density distribution of the gas, three atomic images should be taken, respectively, as $P_{\textrm{abs}}$, $P_{\textrm{ref}}$ and $P_{\textrm{bg}}$. $P_{\textrm{abs}}$ is the absorption image with atom and light, $P_{\textrm{ref}}$ is the reference image without atom, and $P_{\textrm{bg}}$ is the background image without light and atom. The optical density (OD) distribution is obtained from
\begin{equation} \label{49}
P_{\textrm{od}}=\ln\left(\frac{P_{\textrm{ref}}-P_{\textrm{bg}}}{P_{\textrm{abs}}-P_{\textrm{bg}}}\right).
\end{equation}

Due to changes in the intensity and spatial position of the imaging light, $P_{\textrm{ref}}$ is different from the background of $P_{\textrm{abs}}$, leading to fringes and other noises in $P_{\textrm{od}}$ (see Fig. \ref{Sfig4}(a)). As known in Ref. \cite{Ockeloen2010}, these noises can be reduced by using an algorithm to synthesize a new reference image $P_{\textrm{refn}}$, which is closest to the background of $P_{\textrm{abs}}$. Replacing $P_{\textrm{ref}}$ with $P_{\textrm{refn}}$ in Eq. (\ref{49}), we can optimize the atomic image $P_{\textrm{od}}$.

A set of reference images compose a background library $R$, whose linear superposition gives $P_{\textrm{refn}}$ by
\begin{equation} \label{50}
P_{\textrm{refn}}=RC.
\end{equation}

\noindent The coefficient matrix $C$ is determined by setting the least square difference between $P_{\textrm{refn}}$ and $P_{\textrm{abs}}$ in the regions without absorption of atoms. Setting the partial derivative of the square difference with respect to the coefficient to zero, a set of equations are obtained, where the solutions give the coefficients.

Fig. \ref{Sfig4} displays an example of the image optimization. Without optimization, there are many fringes and noises in the background, as shown in Fig. \ref{Sfig4}(a). After optimization, the fringes and noises are greatly reduced in Fig. \ref{Sfig4}(b). In Fig. \ref{Sfig4}(c), we use a Gaussian function to fit the one dimensional OD,
\begin{equation} \label{Gaussian}
\textrm{OD}(x)=\textrm{OD}_{0}+A\exp\left(\frac{-x^{2}}{2\sigma^{2}_{x}}\right).
\end{equation}

\noindent Without optimization, $\textrm{OD}_{0}=0.63\pm0.13$, $\sigma_{x}=27.26\pm0.87$, $\textrm{A}=10.73\pm0.27$, and $\chi^{2}$ of the fitting is 0.45. After optimization, $\textrm{OD}_{0}=0.20\pm0.09$, $\sigma_{x}=25.95\pm0.62$, $\textrm{A}=10.93\pm0.20$, and $\chi^{2}$ of the fitting is 0.26. It can be seen that, through image optimization, the background level and fitting uncertainty are well reduced. The atomic cloud radius also changes, which should be more accurate.

\begin{figure}[htbp]
\centerline{\includegraphics[width=17cm]{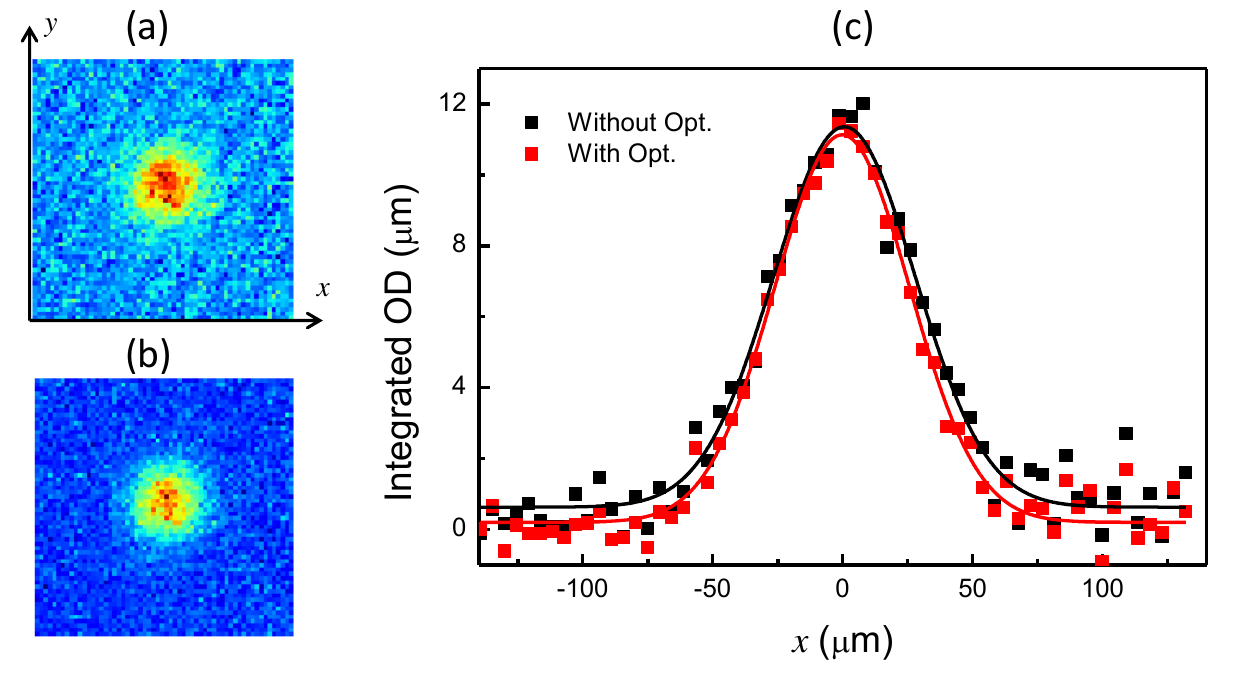}}
\caption{Optimization of atomic images. (a) is for the atomic image without optimization and (b) with optimization. The integrated OD in $x$ direction is shown in (c). The solid line is the fitting using a Gaussian function. The OD distributions with and without optimization are displayed together for comparison. }  \label{Sfig4}
\end{figure}

\section{Hydrodynamic description of the atomic expansion from an isotropic trap}
A Gaussian distribution is used to fit the atomic density profile, $n(x)=\textrm{A} e^{(-x^2/2\sigma_x^2)}$, and the fitted value of $\sigma_x$ is related to the mean square cloud size in $x$ direction by
\begin{equation} \label{26}
{\langle}x^2{\rangle}=\frac{1}{N}\int{nx^2\text{d}x}=\sigma_x^2.
\end{equation}
We consider how ${\langle}x_i^2{\rangle}$ ($i = x, y, z$) evolves with expansion time $t$,
\begin{equation} \label{27}
\begin{aligned}
\frac{\text{d}{\langle}x_i^2{\rangle}}{\text{d}t}&=\frac{1}{N}\int{\frac
{\partial n}{\partial t}x_i^2\text{d}x}.
\end{aligned}
\end{equation}
Using the continuity equation of the hydrodynamic description for one component fluid \cite{Cao2011, Cao2011a, houScalingSolutionsTwofluid2013}, ${\partial n}/{\partial t}+\nabla\cdot(n \mathbf{v})=0$, Eq. (\ref{27}) can be written as
\begin{equation} \label{28}
\begin{aligned}
\frac{\text{d}{\langle}x_i^2{\rangle}}{\text{d}t} & =
\frac{1}{N}\int{[-\nabla\cdot(n \mathbf{v})]x_i^2\text{d}x} \\ & =
\frac{1}{N}\int{[-\nabla\cdot(x_i^2)]n \mathbf{v}\text{d}x}+
\frac{1}{N}\int{[-\nabla\cdot(x_i^2n \mathbf{v})]\text{d}x} \\
& =\frac{1}{N}\int{[-\nabla\cdot(x_i^2)]n \mathbf{v}\text{d}x} \\
& =\frac{1}{N}\int{2x_i n {v_i}\text{d}x}\\
&=2{\langle}x_i{v_i}{\rangle}.
\end{aligned}
\end{equation}
Using the same procedure, we can write the evolution of ${\langle}x_i{v_i}{\rangle}$ with time:
\begin{equation} \label{29}
\begin{aligned}
\frac{\text{d}{\langle}x_iv_i{\rangle}}{\text{d}t} &
= \frac{1}{N}\int{\frac{\partial n}{\partial t}x_iv_i\text{d}x}
+ \frac{1}{N}\int{nx_i\frac{\partial v_i}{\partial t}\text{d}x} \\ &
=\frac{1}{N}\int{[\nabla\cdot(x_iv_i)]n \mathbf{v}\text{d}x}+
\frac{1}{N}\int{nx_i\frac{\partial v_i}{\partial t}\text{d}x} \\ &
={\langle}x(\partial _t+\mathbf{v}\cdot\nabla)v_i{\rangle}
+{\langle}v_i^2{\rangle}.
 \end{aligned}
\end{equation}
Combining Eq. (\ref{28}) and (\ref{29}), we obtain
\begin{equation} \label{30}
\begin{aligned}
\frac{\text{d}^2}{\text{dt}^2 }\frac{{\langle}x_i^2{\rangle}}{2} &
={\langle}x_i(\partial _t+\mathbf{v}\cdot\nabla)v_i{\rangle}
+{\langle}v_i^2{\rangle}.
 \end{aligned}
\end{equation}
The first term is similar to Euler's equation,
\begin{equation} \label{31}
mn(\partial _t+\mathbf{v}\cdot\nabla)v
=-\partial _i
P-n\partial _iU+\sum_j\partial _j(\eta\sigma_
{ij}+\zeta\sigma'\delta_{ij}),
\end{equation}
where $m$ is the atomic mass, $P$ is the scalar pressure, $U$ is the trapping potential energy, and the last term on the right side denotes the friction forces due to shear $\eta$ and bulk $\zeta$ viscosity. The viscosity can be written as $\eta \equiv \alpha_S \hbar n$ and $\zeta \equiv\alpha_B \hbar n$. $\sigma_{ij} \equiv {\partial v_i}/{\partial x_i}+{\partial v_j}/{\partial x_j}-2/3\delta_{ij}\nabla\cdot\mathbf{v}$, $\sigma' =\nabla\cdot\mathbf{v}$. We then take the density averaged product of the Euler's equation with a position component, Eq. (\ref{31}) can be written as
\begin{equation} \label{32}
\begin{aligned}
\frac{1}{N}\int nx_i (\partial _t+\mathbf{v}\cdot\nabla)v_i \text{d}
^3r
&=-\frac{1}{Nm}\int x_i \partial _iP \text{d}^3r
-\frac{1}{Nm}\int x_i n\partial _iU \text{d}^3r
+\frac{1}{Nm}\sum_j\int x_i\partial _j(\eta\sigma_
{ij}+\zeta\sigma'\delta_{ij})\text{d}^3r \\
&=\frac{1}{Nm}\int P \text{d}^3r
-\frac{1}{Nm}\int x_i n\partial _iU \text{d}^3r
-\frac{1}{Nm}\int (\eta\sigma_{ii}
+\zeta\sigma')\text{d}^3r.
\end{aligned}
\end{equation}
The pressure, trapping potential and viscosity terms must be zero for $x_i\rightarrow\pm\infty$. Then combining Eq. (\ref{30}) and (\ref{32}), we obtain
\begin{equation} \label{33}
\frac{\text{d}^2}{\text{dt}^2}\frac{{\langle}x_i^2{\rangle}}{2}
=\frac{1}{Nm}\int P \text{d}^3r
-\frac{1}{m}{\langle} x_i\partial _iU {\rangle}
-\frac{ \hbar}{m}{\langle} \alpha_S\sigma_{ii}
+ \alpha_B\sigma'{\rangle}
+{\langle}v_i^2{\rangle}.
\end{equation}
Eq. (\ref{33}) determines the evolution of the mean square cloud radius, which depends on the conservative forces from the scale pressure and the trapping potential, as well as the dissipative forces arising from the shear and bulk viscosity. For a spherical system, expansion behaviors in all directions are the same, i.e., ${\partial v_i}/{\partial x_i}
={\partial v_j}/{\partial x_j}={\partial v_k}/{\partial x_k}$. Then
\begin{equation} \label{34}
\begin{aligned}
\sigma_{ii} &\equiv \frac{\partial v_i}{\partial x_i}
+\frac{\partial v_i}{\partial x_i}
-\frac{2}{3}\delta_{ii}\nabla\cdot\mathbf{v} \\
& =\frac{\partial v_i}{\partial x_i}
+\frac{\partial v_i}{\partial x_i}
-\frac{2}{3}(\frac{\partial v_i}{\partial x_i}
+\frac{\partial v_j}{\partial x_j}+\frac{\partial v_k}{\partial x_k})
\\ & =0, \\
\sigma'& =\nabla\cdot\mathbf{v}
 = \frac{\partial v_i}{\partial x_i}
+\frac{\partial v_j}{\partial x_j}+\frac{\partial v_k}{\partial x_k}
=3\frac{\partial v_i}{\partial x_i}.
\end{aligned}
\end{equation}
For simplicity, we only consider the atomic expansion in $x$ direction. Inserting Eq. (\ref{34}) into Eq. (\ref{33}), we can find that for a spherical system, the evolution of the mean square cloud radius can be written as
\begin{equation} \label{35}
\frac{\text{d}^2}{\text{dt}^2}\frac{{\langle}x^2{\rangle}}{2}
=\frac{1}{Nm}\int P \text{d}^3r
-\frac{1}{m}{\langle} x\partial _xU {\rangle}
-\frac{ \hbar}{m}{\langle}
3\alpha_B\partial_x{v}_x{\rangle}
+{\langle}v_x^2{\rangle}.
\end{equation}
Now we need to eliminate ${\langle}v_x^2{\rangle}$ by using the energy conservation equation:
\begin{equation} \label{36}
\frac{\text{d}}{\text{d}t}\int \text{d}^3r
(n\frac{1}{2}m\mathbf{v}^2+\varepsilon+nU) =0.
\end{equation}
At $t\geqslant0^+$, atoms are released from the optical trap ($U=0$),
\begin{equation} \label{37}
\begin{aligned}
t=0^+ \qquad E &=\frac{1}{N}\int \text{d}^3r \varepsilon_0, \\
t>0^+
\qquad E & =\frac{1}{N}\int \text{d}^3r\varepsilon+\frac{m}{2}{\langle}\mathbf{v}^2{\rangle}.
\end{aligned}
\end{equation}
Taking $\triangle P\equiv P-{2}/{3}\varepsilon$ into Eq. (\ref{37}),
\begin{equation} \label{38}
\begin{aligned}
\frac{1}{N}\int \text{d}^3r\varepsilon_0 & =
\frac{3}{2N}\int \text{d}^3rP_0 -
\frac{3}{2N}\int \text{d}^3r\triangle P_0 \\ & =
\frac{1}{N}\int \text{d}^3r
\varepsilon+\frac{m}{2}{\langle}\mathbf{v}^2{\rangle}.
\end{aligned}
\end{equation}
Then we find
\begin{equation} \label{39}
\begin{aligned}
\langle v^2 \rangle & =\frac{2}{m}
(\frac{3}{2N}\int \text{d}^3rP_0 -
\frac{3}{2N}\int \text{d}^3r\triangle P_0 -
\frac{1}{N}\int \text{d}^3r\varepsilon)\\
& =\frac{3}{m}\langle r\cdot\nabla U \rangle_0 -
\frac{3}{mN}\int \text{d}^3r\triangle P_0 -
\frac{2}{mN}\int \text{d}^3r\varepsilon .
\end{aligned}
\end{equation}
For a spherical system, $\langle v_x^2 \rangle=\langle v^2 \rangle/3$.

Before release from the trap at $t=0^-$, $\mathbf{v}=0$. Eq. (\ref{35}) can be written as
\begin{equation} \label{40}
\frac{1}{N}\int \text{d}^3r P_0 =\langle {x}\partial_x U\rangle_0,
\end{equation}
where the subscript $()_0$ describe the initial condition in the trap. Combining Eqs. (\ref{39}), (\ref{40}) and (\ref{35}), we obtain
\begin{equation} \label{41}
\begin{aligned}
\frac{\text{d}^2}{\text{d}^2 t}\frac{m{\langle}x^2{\rangle}}{2}
& =\langle x\partial_x U\rangle_0 -
\frac{1}{N}\int \text{d}^3r\triangle P_0 -
\frac{2}{3N}\int \text{d}^3r\varepsilon  \\ &
 +\frac{1}{N}\int \text{d}^3r\triangle P +
\frac{2}{3N}\int \text{d}^3r\varepsilon
- \hbar{\langle} 3\alpha_B\partial_x{v_x}{\rangle} \\
&=\langle x\partial_x U\rangle_0 +
\frac{1}{N}\int \text{d}^3r (\triangle P-\triangle P_0)
- \hbar{\langle} 3\alpha_B\partial_x{v_x}{\rangle}.
\end {aligned}
\end{equation}
This is the atomic expansion evolution from an isotropic harmonic trap. For a unitary Fermi gas, $\triangle P=0$ and $\alpha_B=0$. Eq. (\ref{41}) can be written as
\begin{equation} \label{42}
\begin{split}
\frac{d^2}{dt^2}\frac{m{\langle}x^2{\rangle}}{2}={\langle}x
{\frac{\partial U}{\partial x}}{\rangle}_0,
\end{split}
\end{equation}

\begin{equation} \label{43}
{\langle}x^2{\rangle}={\langle}x^2{\rangle}_0+ \omega_x^2{\langle}x^2{\rangle}_0t^2.
\end{equation}
Eqs. (\ref{42}) and (\ref{43}) display the scale-invariant expansion of a unitary Fermi gas from an isotropic trap, which is similar to a non-interacting Fermi gas. While away from the resonant interaction, $\triangle P\neq0$ and $\alpha_B\neq0$, the scale-invariant expansion will be broken.

\section{Interaction effect on the ballistic expansion}
For our experiment with a unitary Fermi gas in a spherical trap, the shear viscosity contribution vanishes, and the bulk viscosity is zero. In this condition, we could not measure the viscosity from the expansion behaviors. But we could still observe the interaction effects on the ballistic expansion. Though $\tau^2(t)$ shows no difference between the interacting and noninteracting gasses, the absolute size ${\langle}x^2{\rangle}$ during the expansion is dependent on the interaction, as seen in Eq. (2) of the main text. The interaction is included in the in-situ atomic cloud size ${\langle}x_{0}^2{\rangle}$. In Fig. \ref{Sfig5}, we explicitly demonstrate the interaction effects by comparing the theoretical calculation and experimental results. We calculate the expansion behavior according to Eq. (2) of the main text with the experimental parameters (atomic temperature, atom number and trapping frequency). The expansion behavior observed in the experiment agrees well with the calculation including interaction, while deviates obviously from the calculation without interaction. The absolute size ${\langle}x^2{\rangle}$ without interaction is bigger than that with interaction.

\begin{figure}[htbp]
\centerline{\includegraphics[width=17cm]{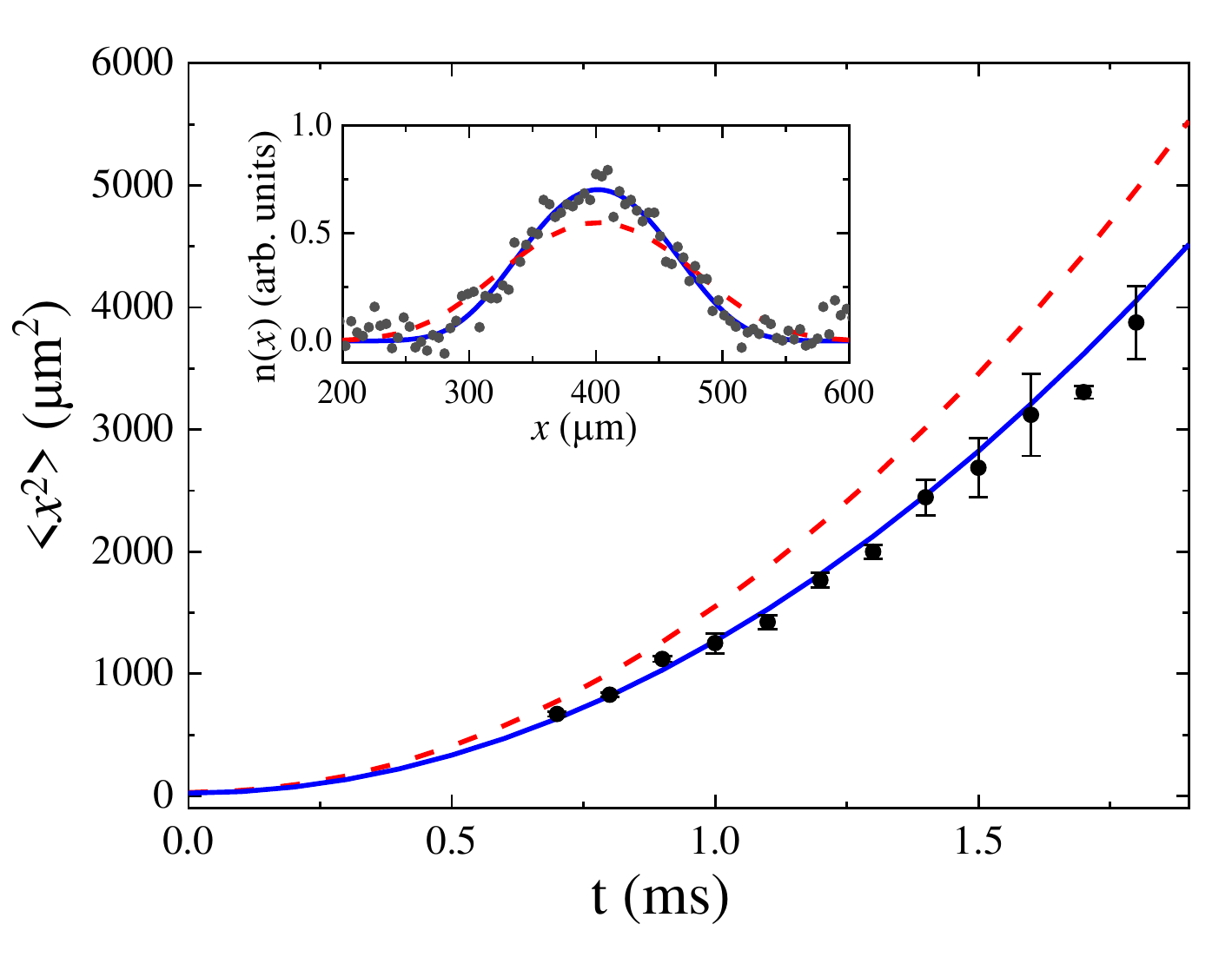}}
\caption{Expansion of a unitary Fermi gas ($a\rightarrow\infty$) and an ideal Fermi gas ($a\rightarrow 0$). The black circles are for the experiments at $a\rightarrow\infty$, the blue solid (red dashed) curve is for the theoretical calculation at $a\rightarrow\infty$ ($a\rightarrow0$). The atomic temperature is $T/T_{F}=0.3$, the atom number is $N=2.3\times10^{4}$, and the trapping frequency is $\omega_{0} = 2\pi\times 1200 \ \textrm{Hz}$. Inset: 1D density distribution of atomic cloud after an expansion time of 1.8 ms, with the blue solid (red dashed) curve denoting the theoretical result at $a\rightarrow\infty$ ($a\rightarrow 0$). }  \label{Sfig5}
\end{figure}

\section{SO(2,1) Symmetry}
The unitary Fermi gas confined in an isotropic harmonic trap has the SO(2,1) symmetry, which has been introduced with details in Reference \cite{Werner2006}. A spherically trapped unitary Fermi gas of $N$ atoms is described by the Hamiltonian
\begin{align}
H=\sum_{j=1}^{N}\left(-\frac{\hbar^2}{2m}\nabla^2_j+\frac{1}{2}m\omega^2_0r^2_j\right),
\end{align}
where the interaction between atoms could alternatively be characterized by a short-range boundary condition obeyed by the wave function $\psi(r\rightarrow 0) \sim 1/r$. Let us introduce raising/lowing operators,
\begin{align}
L_+=+\frac{3N}{2}+\hat{D}+\frac{H}{\hbar\omega_0}-\frac{m\omega_0}{\hbar}X^2,\\
L_-=-\frac{3N}{2}-\hat{D}+\frac{H}{\hbar\omega_0}-\frac{m\omega_0}{\hbar}X^2,
\end{align}
\noindent where $\hat{D}\equiv \vec{X}\cdot \partial _{\vec{X}}$ and $X^2=\sum_{j=1}^{N}r^2_j$. Repeated action of $L_+$ and $L_-$ on an eigenstate $\psi$ with energy $E$ will thus generate a ladder of eigenstates with a regular energy $E\pm2\hbar\omega_0$.  It is found that
\begin{align}
[H,L_+]=2\hbar\omega_0 L_+,\\
[H,L_-]=-2\hbar\omega_0 L_-,\\
[L_+,L_-]=-4\frac{H}{\hbar\omega_0},
\end{align}
which satisfies the algebra of SO(2,1). The symmetry is SO(2,1), but not SO(3), because of the minus sign appearing in the last equality \cite{PitaevskiiPRA1997Scale}.

Remarkably, action $L_+$ on the lowest energy $|\psi_0\rangle$ with energy $E_0$, we find
\begin{align}
HL_+|\psi_0\rangle=(E_0+2\hbar\omega_0)L_+|\psi_0\rangle.
\end{align}
Then $L_+|\psi_0\rangle$ is an eigenstate with energy $E_0+2\hbar\omega_0$. This corresponds to the lowest breathing mode with the oscillation frequency $\omega_B=2\omega_0$.

From the general theory of Lie algebras, one may form the so-called Casimir operator,
\begin{align}
\hat{\mathcal{C}}=H^2-\frac{1}{2}(\hbar\omega_0)^2(L_+L_-+L_-L_+),
\end{align}
which commutes with $H$ and $L_{\pm}$. So, $\hat{\mathcal{C}}$ is a scalar within each ladder. For the ground energy ladder, we have
\begin{align}
\hat{\mathcal{C}}|\psi_0\rangle=E_0\left(E_0-2\hbar\omega_0\right)|\psi_0\rangle.
\end{align}

\end{widetext}


\begin{thebibliography}{10}

\bibitem{Chin2010}
C.~Chin, R.~Grimm, P.~Julienne, and E.~Tiesinga.
\newblock Feshbach resonances in ultracold gases.
\newblock {Rev. Mod. Phys.} 82, 1225 (2010).

\bibitem{OHara2002}
K.~M. O'Hara, S.~L. Hemmer, M.~E. Gehm, S.~R. Granade, and J.~E. Thomas.
\newblock Observation of a strongly interacting degenerate {Fermi} gas of
  atoms.
\newblock {Science} 298, 2179 (2002).

\bibitem{Kinast2005}
J.~Kinast, A.~Turlapov, J.~E. Thomas, Q.~Chen, J.~Stajic, and K.~Levin.
\newblock Heat capacity of a strongly interacting {Fermi} gas.
\newblock {Science} 307, 1296 (2005).

\bibitem{Nascimbene2010}
S.~Nascimb{\`e}ne, N.~Navon, K.~J. Jiang, F.~Chevy, and C.~Salomon.
\newblock Exploring the thermodynamics of a universal {Fermi} gas.
\newblock {Nature} 463, 1057 (2010).

\bibitem{Ku2012}
M.~J.~H. Ku, A.~T. Sommer, L.~W. Cheuk, and M.~W. Zwierlein.
\newblock Revealing the superfluid lambda transition in the universal
  thermodynamics of a unitary {Fermi} gas.
\newblock {Science} 335, 563 (2012).

\bibitem{Sidorenkov2013}
L.~A. Sidorenkov, M.~K. Tey, R.~Grimm, Y.-H. Hou, L.~Pitaevskii, and
  S.~Stringari.
\newblock Second sound and the superfluid fraction in a {Fermi} gas with
  resonant interactions.
\newblock {Nature} 498, 78 (2013).

\bibitem{Bardon2014}
A.~B. Bardon, S.~Beattie, C.~Luciuk, W.~Cairncross, D.~Fine, N.~S. Cheng,
  G.~J.~A. Edge, E.~Taylor, S.~Zhang, S.~Trotzky, and J.~H. Thywissen.
\newblock Transverse demagnetization dynamics of a unitary {Fermi} gas.
\newblock {Science} 344, 722 (2014).

\bibitem{Mukherjee2017}
B.~Mukherjee, Z.~Yan, P.~B. Patel, Z.~Hadzibabic, T.~Yefsah, J.~Struck and M.~W.
  Zwierlein.
\newblock Homogeneous atomic Fermi gases.
\newblock {Phys. Rev. Lett.} 118, 123401 (2017).

\bibitem{Patel2020}
P.~B. Patel, Z.~Yan, B.~Mukherjee, R.~J. Fletcher, J.~Struck, and M.~W.
  Zwierlein.
\newblock Universal sound diffusion in a strongly interacting {Fermi} gas.
\newblock {Science} 370, 1222 (2020).

\bibitem{Li2022}
X.~Li, X.~Luo, S.~Wang, K.~Xie, X.-P. Liu, H.~Hu, Y.-A. Chen, X.-C. Yao, and
  J.-W. Pan.
\newblock Second sound attenuation near quantum criticality.
\newblock {Science} 375, 528 (2022).

\bibitem{Yan2024}
Z.~Yan, P.~B. Patel, B.~Mukherjee, C.~J. Vale, R.~J. Fletcher, and M.~W.
  Zwierlein.
\newblock Thermography of the superfluid transition in a strongly interacting Fermi gas.
\newblock {Science} 383, 629 (2024).

\bibitem{Menotti2002}
C.~Menotti, P.~Pedri, and S.~Stringari.
\newblock Expansion of an interacting {Fermi} gas.
\newblock {Phys. Rev. Lett.} 89, 250402 (2002).

\bibitem{Thomas2005}
J.~E. Thomas, J.~Kinast, and A.~Turlapov.
\newblock Virial theorem and universality in a unitary {Fermi} gas.
\newblock {Phys. Rev. Lett.} 95, 120402 (2005).

\bibitem{Elliott2014}
E.~Elliott, J.~A. Joseph, and J.~E. Thomas.
\newblock Observation of conformal symmetry breaking and scale invariance in
  expanding {Fermi} gases.
\newblock {Phys. Rev. Lett.} 112, 040405 (2014).

\bibitem{Deng2016}
S. Deng, Z.-Y. Shi, P.~Diao, Q.~Yu, H.~Zhai, R.~Qi, and H.~Wu.
\newblock Observation of the efimovian expansion in scale-invariant {Fermi}
  gases.
\newblock {Science} 353, 371 (2016).

\bibitem{Cao2011a}
C.~Cao, E.~Elliott, H.~Wu, and J.~E. Thomas.
\newblock Searching for perfect fluids: Quantum viscosity in a universal
  {Fermi} gas.
\newblock {New J. Phys.} 13, 075007 (2011).

\bibitem{Levin2011viscosity}
H.~Guo, D.~Wulin, C.-C. Chien, and K.~Levin.
\newblock Microscopic approach to shear viscosities of unitary {Fermi} gases
  above and below the superfluid transition.
\newblock {Phys. Rev. Lett.} 107, 020403 (2011).

\bibitem{Shafer2017viscosity}
M.~Bluhm, J.~Hou, and T.~Sch\"afer.
\newblock Determination of the density and temperature dependence of the shear
  viscosity of a unitary {Fermi} gas based on hydrodynamic flow.
\newblock {Phys. Rev. Lett.} 119, 065302 (2017).

\bibitem{Cao2011}
C.~Cao, E.~Elliott, J.~Joseph, H.~Wu, J.~Petricka, T.~Sch{\"a}fer, and J.~E.
  Thomas.
\newblock Universal quantum viscosity in a unitary {Fermi} gas.
\newblock {Science} 331, 58 (2011).

\bibitem{Thomas2014anomalousViscosity}
E.~Elliott, J.~A. Joseph, and J.~E. Thomas.
\newblock Anomalous minimum in the shear viscosity of a {Fermi} gas.
\newblock {Phys. Rev. Lett.} 113, 020406 (2014).

\bibitem{Thomas2015superfluidViscosity}
J.~A. Joseph, E.~Elliott, and J.~E. Thomas.
\newblock Shear viscosity of a unitary {Fermi} gas near the superfluid phase
  transition.
\newblock {Phys. Rev. Lett.} 115, 020401 (2015).

\bibitem{Werner2006}
F.~Werner and Y.~Castin.
\newblock Unitary gas in an isotropic harmonic trap: Symmetry properties and
  applications.
\newblock {Phys. Rev. A} 74, 053604 (2006).

\bibitem{Supplemental2023}
See {Supplemental Material} for the formation of an isotropic trap, measurement of the trapping frequency, determination of the atomic temperature and in-situ atomic cloud size, fringe-removal analysis of the atomic images, hydrodynamic description of the atomic expansion from an isotropic trap, interaction effect on the ballistic expansion, and SO(2,1) symmetry, which includs Refs. \cite{veeravalliBraggSpectroscopyStrongly2008a, luo2009thermodynamic, houScalingSolutionsTwofluid2013}.

\bibitem{veeravalliBraggSpectroscopyStrongly2008a}
G.~Veeravalli, E.~Kuhnle, P.~Dyke, and C.~J. Vale.
\newblock Bragg spectroscopy of a strongly interacting {Fermi} gas.
\newblock {Phys. Rev. Lett.} 101, 250403 (2008).

\bibitem{luo2009thermodynamic}
L.~Luo and J.~E. Thomas.
\newblock Thermodynamic measurements in a strongly interacting fermi gas.
\newblock {J. Low Temp. Phys.} 154, 1 (2009).

\bibitem{houScalingSolutionsTwofluid2013}
Y.~Hou, L.~P. Pitaevskii, and S.~Stringari.
\newblock Scaling solutions of the two-fluid hydrodynamic equations in a
  harmonically trapped gas at unitarity.
\newblock {Phys. Rev. A} 87, 033620 (2013).

\bibitem{Ho2004}
T.-L. Ho.
\newblock Universal thermodynamics of degenerate quantum gases in the unitarity
  limit.
\newblock {Phys. Rev. Lett.} 92, 090402 (2004).

\bibitem{Son2007}
D.~T. Son.
\newblock Vanishing bulk viscosities and conformal invariance of the unitary
  {Fermi} gas.
\newblock {Phys. Rev. Lett.} 98, 020604 (2007).

\bibitem{Escobedo2009}
M.~A. Escobedo, M.~Mannarelli, and C.~Manuel.
\newblock Bulk viscosities for cold {Fermi} superfluids close to the unitary
  limit.
\newblock {Phys. Rev. A} 79, 063623 (2009).

\bibitem{Dusling2013}
K.~Dusling and T.~Sch{\"a}fer.
\newblock Bulk viscosity and conformal symmetry breaking in the dilute {Fermi}
  gas near unitarity.
\newblock {Phys. Rev. Lett.} 111, 120603 (2013).

\bibitem{Yan2021}
X.~Yan, D.~Sun, L.~Wang, J.~Min, S.~Peng, and K.~Jiang.
\newblock Production of degenerate {Fermi} gases of \textsuperscript{6}{Li}
  atoms in an optical dipole trap.
\newblock {Chin. Phys. Lett.} 38, 056701 (2021).

\bibitem{Yan2022}
X.~Yan, D.~Sun, L.~Wang, J.~Min, S.~Peng, and K.~Jiang.
\newblock Observation of the {BEC-BCS} crossover in a degenerate {Fermi} gas of
  lithium atoms.
\newblock {Chin. Phys. B} 31, 016701 (2022).

\bibitem{Ockeloen2010}
C.~F. Ockeloen, A.~F. Tauschinsky, R.~J.~C. Spreeuw, and S.~Whitlock.
\newblock Detection of small atom numbers through image processing.
\newblock {Phys. Rev. A} 82, 061606(R) (2010).

\bibitem{Stringari1999RMP}
F.~Dalfovo, S.~Giorgini, L.~P. Pitaevskii, and S.~Stringari.
\newblock Theory of {Bose-Einstein} condensation in trapped gases.
\newblock {Rev. Mod. Phys.} 71, 463 (1999).

\bibitem{Cooper1997PRL}
M.~J. Holland, D.~S. Jin, M.~L. Chiofalo, and J.~Cooper.
\newblock Emergence of interaction effects in {Bose-Einstein} condensation.
\newblock {Phys. Rev. Lett.} 78, 3801 (1997).

\bibitem{liExpansionDynamicsSpherical2019}
R.~Li, T.~Gao, D.~Zhang, S.~Peng, L.~Kong, X.~Shen, and K.~Jiang.
\newblock Expansion dynamics of a spherical {{Bose}}\textendash{{Einstein}}
  condensate.
\newblock {Chin. Phys. B} 28, 106701 (2019).

\bibitem{PitaevskiiPRA1997Scale}
L. P. Pitaevskii and A. Rosch.
\newblock Breathing modes and hidden symmetry of trapped atoms in two dimensions.
\newblock {Phys. Rev. A} 55, R853 (1997).

\bibitem{DalibardPRL2002Transverse}
F. Chevy, V. Bretin, P. Rosenbusch, K. W. Madison and J. Dalibard.
\newblock Transverse Breathing Mode of an Elongated Bose-Einstein Condensate.
\newblock { Phys. Rev. Lett.} 88, 250402 (2002).

\bibitem{VogtPRL2012Scale}
E.~Vogt, M.~Feld, B.~Fr$\ddot{\textrm{o}}$hlich, D.~Pertot, M.~Koschorreck, and M.~K$\ddot{\textrm{o}}$hl.
\newblock Scale Invariance and Viscosity of a Two-Dimensional Fermi Gas.
\newblock {Phys. Rev. Lett.} 108, 070404 (2012).

\bibitem{Stringari2008RMP}
S.~Giorgini, L.~P. Pitaevskii, and S.~Stringari.
\newblock Theory of ultracold atomic {Fermi} gases.
\newblock {Rev. Mod. Phys.} 80, 1215 (2008).

\bibitem{huCollectiveModesBallistic2004}
H.~Hu, A.~Minguzzi, X.-J. Liu, and M.~P. Tosi.
\newblock Collective modes and ballistic expansion of a {{Fermi}} gas in the
  {{BCS-BEC}} crossover.
\newblock {Phys. Rev. Lett.} 93, 190403 (2004).

\bibitem{Heiselberg2004PRL}
H.~Heiselberg.
\newblock Collective modes of trapped gases at the {BEC-BCS} crossover.
\newblock {Phys. Rev. Lett.} 93, 040402 (2004).

\bibitem{wangOscillatorylikeExpansionFermionic2020}
X. Wang, Y. Wu, X. Liu, Y. Wang, H. Chen, M.~Maraj, Y.~Deng,
  X.-C. Yao, Y.-A. Chen, and J.-W. Pan.
\newblock Oscillatory-like expansion of a fermionic superfluid.
\newblock {Sci. Bull.} 65, 7 (2020).

\bibitem{Navon2010}
V.~Navon, S.~Nascimb$\grave{\textrm{e}}$ne, F.~Chevy, and C.~Salomon.
\newblock The Equation of State of a Low-Temperature Fermi Gas with Tunable Interactions.
\newblock {Science} 328, 5979 (2010).

\bibitem{KinastBreakdown2004PRA}
J.~Kinast, A.~Turlapov, and J.~E. Thomas.
\newblock Breakdown of hydrodynamics in the radial breathing mode of a strongly interacting Fermi gas.
\newblock {Phys. Rev. A.} 70, 051401(R) (2004).

\bibitem{Zhou2020PRLgeometic}
C. Lyu, C. Lv and Q. Zhou.
\newblock Geometrizing quantum dynamics of a {Bose-Einstein} condensate.
\newblock { Phys. Rev. Lett.} 125, 253401 (2020).

\bibitem{Zhou2020PRLbreather}
C. Lv, R. Zhang, Q. Zhou,
\newblock SU(1,1) echoes for breathers in quantum Gases.
\newblock { Phys. Rev. Lett.} 125, 253002 (2020).

\bibitem{ZhoufeiPRA2019Conformal}
J.~Maki and F.~Zhou.
\newblock Quantum many-body conformal dynamics: Symmetries, geometry, conformal
  tower states, and entropy production.
\newblock {Phys. Rev. A} 100, 023601 (2019).

\bibitem{ZhoufeiPRA2020Conformal}
J.~Maki and F.~Zhou.
\newblock Far-away-from-equilibrium quantum-critical conformal dynamics:
  Reversibility, thermalization, and hydrodynamics.
\newblock {Phys. Rev. A} 102, 063319 (2020).

\bibitem{ZhoufeiPRL2022Conformal}
J.~Maki, S.~Zhang, and F.~Zhou.
\newblock Dynamics of strongly interacting {Fermi} gases with time-dependent
  interactions: Consequence of conformal symmetry.
\newblock {Phys. Rev. Lett.} 128, 040401 (2022).

\bibitem{Enss2019}
T. ~Enss.
\newblock Bulk viscosity and contact correlations in attractive {Fermi} gases.
\newblock {Phys. Rev. Lett.} 123, 205301 (2019).

\bibitem{Hofmann2020PRAviscosity}
J. ~Hofmann.
\newblock High-temperature expansion of the viscosity in interacting quantum gases.
\newblock {Phys. Rev. A} 101, 013620 (2020).

\bibitem{Nishida2019AnnalsVisocosity}
Y. ~Nishida.
\newblock Viscosity spectral functions of resonating fermions in the quantum virial expansion.
\newblock {Ann. Phys.} 410, 167949 (2019).


\end{thebibliography}
\end{document}